\documentclass[fleqn,usenatbib]{mnras}
\usepackage{hyperref}
\usepackage[T1]{fontenc}
\usepackage{ae,aecompl}
\usepackage{graphicx}
\usepackage{enumerate}
\usepackage{amsmath}
\let\oldAA\AA
\renewcommand{\AA}{\text{\normalfont\oldAA}}
\usepackage{array}
\usepackage{makecell}
\usepackage{amssymb}
\usepackage{booktabs}
\usepackage{pdflscape} 
\usepackage{footnote}
\usepackage{threeparttable}
\usepackage{epsfig}
\usepackage{pslatex}
\usepackage{subfig}
\usepackage{comment}
\usepackage{epstopdf}
\usepackage{csquotes}
\usepackage{multirow}
\usepackage{orcidlink}
\usepackage{soul}
\graphicspath{{./}{figures/}}
\newcommand{\ergscmA}{\,erg\,s$^{-1}$\,cm$^{-2}$\,\AA$^{-1}$}
\newcommand{\kms}{\,km\,s$^{-1}$}
\newcommand{\ergs}{\,erg\,s$^{-1}$}
\title[Nova Her 2021  (V1674 Her)]{Study of the fastest classical nova, V1674 Her: Photoionization and Morpho-kinemetic model analysis }
\author[]{Gesesew R. Habtie\,\orcidlink{0000-0001-9827-738X}$^{1}$\thanks{E-mail: meetgesese@gmail.com},
Ramkrishna Das\,\orcidlink{0000-0002-5440-7186},$^{1}$
Ruchi Pandey\,\orcidlink{0000-0002-6222-3045},$^{2}$
N. M. Ashok\,$^{2}$, and
\newauthor
Pavol A. Dubovsky$^{3}$
\\
$^{1}$ S N Bose National Centre for Basic Sciences, Salt Lake, Kolkata 700 106, India\\
$^{2}$ Physical Research Laboratory, Ahmedabad, State of Gujarat, India\\
$^{3}$ Vihorlat Observatory in Humenne, Mierova 4, 06601 Humenne, Slovakia
\\
}

\date{Accepted 2023 October 24. Received 2023 October 24; in original form 2023 August 16}

\begin{document}
\label{firstpage}
\pagerange{\pageref{firstpage}--\pageref{lastpage}}
\maketitle
\begin{abstract}
We present the results of the investigation of the nova V1674 Her (2021), recognised as the swiftest classical nova, with $t_2 \sim 0.90$ days. The distance to the nova is estimated to be 4.97 kpc. The mass and radius of the WD are calculated to be $\sim~1.36~M_{\sun}$ and $\sim 0.15~R_{\earth}$, respectively. Over the course of one month following the outburst, V1674 Her traversed distinct phases--pre-maxima, early decline, nebular, and coronal--displaying a remarkably swift transformation. The nebular lines emerged on day 10.00, making it the classical nova with the earliest observed commencement to date.  We modelled the observed optical spectrum using the photoionization code \textsc{cloudy}. From the best-fitting model we deduced different physical and chemical parameters associated withe the system. The temperature and luminosity of the central ionizing sources are found in the range of $1.99 - 2.34~\times 10^5$  K and $1.26 - 3.16~ \times 10^{38}$ \ergs, respectively. Elements such as \ion{He}{}, \ion{O}{}, \ion{N}{}, and \ion{Ne}{} are found to be overabundant compared to solar abundance in both the nebular and coronal phases. According to the model, Fe II abundance diminishes while Ne abundance increases, potentially elucidating the rare hybrid transition between Fe and He/N nova classes. The ejected mass across all epochs spanned from $3.42  - 7.04~ \times 10^{-5}~M_{\sun}$. Morpho-kinematic modelling utilising \textsc{shape} revealed that the nova V1674 Her possesses a bipolar structure with an equatorial ring at the centre and an inclination angle of i = 67$\pm$ 1.5$^{\circ}$.
\end{abstract}


\section{Introduction}\label{sec1}
Classical novae (CNe) are interacting binary systems consisting of a CO or ONe-type white dwarf (WD) primary and a Roche-lobe-filling late-type secondary. Over an extended period, the WD accretes hydrogen-rich material from the secondary and accumulates it onto its surface at a rate of $\dot{M} \sim 10^{-9} - 10^{-10}~M_{\sun}$ yr$^{-1}$ \citep{2007ApJJose}. It forms a degenerate fuel layer until the thermonuclear runaway (TNR) process is triggered, which gives birth of Nova explosion \citep{1993prplGehrz, starrfieldiliadishix2008, 2016PASPStarrfield}. As a consequence, an explosion occurs on the surface of the WD, ejecting a significant amount of mass ($\sim 10^{-4} - 10^{-5}$ M$_{\sun}$) from the system during an outburst \citep{ 1998PASPGehrz, 2020ApJStarrfield}. The ejecta of a nova contains different heavy and light metals (such as \ion{Be}{}, \ion{C}{}, \ion{N}{}, \ion{O}{}, \ion{Mg}{}, and \ion{Fe}{}) \citep{2016PASPStarrfield}, grains (such as carbon and silicate grains) \citep{1980ApJGehrzz, Kyung-Won1994, 1998PASPGehrz}, molecules (such as \ion{H$_2$}{}, \ion{CO}{}, \ion{CN}{}, \ion{PAH}{} etc), and radioactive nuclei (such as $^7$\ion{Be}{}, $^{13}$\ion{N}{}, $^{18}$\ion{F}{}, $^{22}$\ion{Na}{}, and $^{26}$\ion{Al}{}) \citep{1998PASPGehrz, 1998MNRASGomez-Gomar, 2002AIPCHernanz, 2020ApJStarrfield}.

After the explosion, the system's brightness increases to its peak and then gradually declines based on its rate of decline. This rate is defined by the characteristic time, $t_{\text{n}}$, which represents the number of days the nova takes to decline by \lq{n}\rq~ magnitudes from its peak brightness. Classical novae are classified into different speed classes \citep{1964book_Gaposchkin}: very fast ($t_2<10$ days), fast ($t_2=11-25$ days), moderately fast ($t_2=26-80$ days), slow ($t_2=81-150$ days), and very slow ($t_2=151-250$ days).  A few well known examples are: very fast -- V838 Her, 1991 ($t_2=1$ day); fast -- OS And, 1986 ($t_2=11$ days); moderately fast -- V842 Cen, 1986 ($t_2=43$ days); slow -- V1819 Cyg, 1986 ($t_2=95$ days) \citep[][and references therein]{2010AJStrope}.

\section{Nova Her, 2021 (V1674 Her)}
The classical nova \lq{Nova Her 2021}\rq~(V1674 Her) was discovered by Seiji Ueda on  2021 June 12.5484 UT, (JD 2459378.0484), with a visual magnitude of 8.4 at a position $\alpha = 18^h57^m 30^s.95$, $\delta = +16^{\circ}53\arcsec39\arcmin.6$ (J2000) 
(CBET 4976)\footnote{\url{http://www.cbat.eps.harvard.edu/unconf/followups/J18573095+1653396.html}}, which corresponds to the galactic coordinate $l=048^{\circ}.707$, $b= +06^{\circ}.3114$. The nova reached its peak brightness on 2021 June 12.96 UT with a magnitude of V=6.14. The characteristic times ~$t_2,~t_3,~\text{and}~ t_6$ were estimated to be 1 day, 2.2 days, and 14 days, respectively  \citep{2021oodward}. 

The outburst of V1674 Her has been closely monitored across several wavelength domains, from radio to $\gamma$-rays \citep{2021ATel14731Sokolovsky, 2021ApJDrake, 2021ATel14713Vandenbroucke, 2021ATel14710,2021oodward, 2022RNAASWoodward}. \citet{2021ATel14705Li} discovered an uncatalogued gamma-ray source at the nova's position using \textit{Fermi-LAT} data in the range of 0.1-300 GeV. On 2021 June 14.41 UT (day 2.22), Swift observations with the \textit{X-Ray Telescope (XRT)} detected the earliest X-ray emission from V1674 Her, while its super-soft source phase was detected on July 1.09 UT (day 18.9), \citep{2021ATel14747Page, 2021ApJDrake}.

The \ion{H}{$\epsilon$} and \ion{H}{$\zeta$} lines displayed a high-velocity multiple absorption dip reaching $\sim$6500 \kms \citep{2021ATel14736Kuin}.  During the early phases of V1674 Her's evolution, optical and infrared observations revealed the presence of P-Cygni and flat-topped Balmer spectral line profiles. These profiles exhibited extensive substructures and line widths that increased over time, eventually reaching $\sim$11,000 \kms in width \citep[][and references therein]{2021ATel14710, 2021oodward}. 

Radio observations with the \textit{Karl G. Jansky Very Large Array (VLA)} at 2.6, 3.4, 5.1, 7.0, 13.7, 16.5, 31.1, and 34.9 GHz unveiled non-thermal emission \citep{2021ATel14731Sokolovsky}. Multiple observations suggested that the object might be classified as a \enquote{\ion{Fe}{ii} type} nova \citep{2021ATel14723Woodward, 2021ATel14718Albanese, 2021ATel14710, 2021ATelSokolovsky}, primarily due to the presence of several \ion{Fe}{ii} lines. Nevertheless, spectral analysis from observations conducted on the seventeenth day and onwards from the maximum magnitude revealed the emergence of several triply and more ionized neon lines \citep{2021ATel14746, 2022RNAASWoodward}. \citet{2021ATel14746} proposed that this nova belongs to a neon-type subclass of CNe originating from an ONe WD.

In this paper, we study the nova V1674 Her, 2021 during the initial month following its outburst. We model the observed spectra using the photoionization code \textsc{cloudy} (V.17.02 \citep{ferland17}).  From the best-fitting model, we estimate the chemical and physical properties of the nova system, as well as their evolution. The photoionization modelling procedures, results, and discussions are presented in Section \ref{pma} of this paper. We also employ \textsc{shape} \citep{2012Steffen, 2017AC_Steffen}, a morpho-kinematical software, to analyse and dissect the 3D geometry and kinematical structure of the ejecta (Section \ref{mkm}). The results obtained are summarised in Section~\ref{sec7}.

\section{Observations}\label{sec:observation}
In the present study, we used spectroscopic data collected from days -0.032 to +28.8 following the outburst of V1674 Her. The data was acquired from the publicly accessible Astronomical Ring for Access to Spectroscopy Database (ARAS Database\footnote{ \url{https://aras-database.github.io/database/novae.html}}; \citet{2019Teyssier}). The ARAS symbiotic project serves as an archival repository for data and encompasses a network of small telescopes, ranging from 20 to 60 cm in diameter, fitted with spectrographs with resolutions spanning from R $\sim$ 500 to 15,000, covering the wavelength range of 3600 to $\sim$ 9000 {\AA}. In total, a set of 20 optical spectra (ranging from $\sim$ 3700 to 7800 {\AA}) with resolutions varying from 300 to 1200 were employed. These spectra were acquired during the first month subsequent to the eruption and were contributed by various observers from different observatories.
The observational details and the respective observatories are outlined in Table \ref{tab:T1}. The spectroscopic data were processed utilising the Integrated Spectroscopic Innovative Software (\textsc{ISIS}\footnote{\url{http://www.astrosurf.com/buil/isis-software.html}}) following standard procedures. Additionally, we analysed optical \textit{BVRI} photometric data obtained from the American Association of Variable Star Observers (AAVSO)\footnote{\url{https://www.aavso.org/}} database to investigate the photometric behaviour of the nova during the initial month post-outburst.

\begin{table*}
	\caption{Log of optical spectral observation of V1674 Her.   
	}
	\label{tab:T1}
	\centering
	\setlength{\tabcolsep}{5pt}
	\begin{tabular}{l c c c c c c c c c }
		\hline
		MJD&Date 2021 (UT) & t$^a$(days)& Observer &Observatory& Spectrograph& Camera & R$^{b}$ & Coverage (\AA)&TTE$^{c}$(s)\\ 
		\midrule 
		59378.454&June 12.927&-0.032&DBO$^{d}$& WCO$^{i}$&LISA&SXVR-H694&1023&3701-7391&4778\\
		59379.447&June 13.947&0.988 &DBO$^{d}$&WCO$^{i}$&LISA&SXVR-H694& 941&3901-7381&3400\\
		59381.445&June 15.945&2.986 & DBO$^{d}$&WCO$^{i}$&LISA&SXVR-H694&1036&3900-7380&3041\\
		59383.485&June 17.985&5.007 &RLE$^{e}$&THO$^{j}$&ALPY200 lines/mm&ATK428&532&3760-7880&1821 \\
		59385.428&June 19.928&6.969 &PAD$^{f}$&VNT$^{k}$&LISA&Atik 460EX&1027  &3900-7550&3041 \\
		59387.364&June 21.864&8.905 &PAD$^{f}$&VNT$^{k}$&LISA&Atik 460EX&1037 &3900-7550&4867\\
		59388.493&June 22.992 &10.000 &RLE$^{e}$&THO$^{j}$&ALPY200 lines/mm&ATK428&523 &3800-7850&6485\\
		59389.356&June 23.856&10.897 &PAD$^{f}$&VNT$^{k}$&LISA&Atik 460EX&1056&3900-7550&3038  \\
		59391.721&June 26.221&13.230 &FAS$^{g}$&DCO$^{l}$&LISA&Atik414ex&1068&3716-7304&5078  \\
		59393.367&June 27.867&14.908 &PAD$^{f}$&VNT$^{k}$&LISA&Atik 460EX&1122&4000-7550&3038 \\
		59396.402&June 30.901&17.920 &DDJ$^{h}$&LSS$^{m}$&ALPY$_{-}$slit35&Atik428&347&3750-7415&4519 \\
		59396.431&June 30.931&17.972 &PAD$^{f}$&VNT$^{k}$&LISA&Atik 460EX&1149&4100-7550&1810 \\
		59397.408&July 01.908&19.880 &PAD$^{f}$&VNT$^{k}$&LISA&Atik 460EX&1131& 4200-7550&8010 \\
		59400.425&July 04.925&22.920 &DDJ$^{h}$&LSS$^{m}$&ALPY$_{-}$slit35&Atik428&348&3750-7415&4519 \\
		59400.414&July 04.914&22.886 &PAD$^{f}$&VNT$^{k}$&LISA&Atik 460EX&1132 &4150-7550&6018 \\
		59402.411&July 06.910&24.883 &PAD$^{f}$&VNT$^{k}$&LISA&Atik 460EX&1179&4000-7550&6020\\
		59403.366&July 07.866&25.838 &PAD$^{f}$&VNT$^{k}$&LISA&Atik 460EX&1120&4100-7500&6021 \\
		59404.372&July 08.872&26.844&PAD$^{f}$&VNT$^{k}$&LISA&Atik 460EX&1154&4100-7500&4816 \\
		59405.394&July 09.894&27.880 &DDJ$^{h}$&LSS$^{m}$&ALPY$_{-}$slit35&Atik428&347&3762-7531&5425 \\
		59406.452&July 10.952&28.924 &PAD$^{f}$&VNT$^{k}$&LISA&Atik 460EX&1138&4150-7500&4815\\
		\hline
	\end{tabular}\\
	{\raggedright Note: $^{(a)}$Number of days counted from $t_0$ (2021 June 12.959 UT, MJD 59378.459),  $^{(b)}$Resolution, $^{(c)}$Total Time of Exposure,  $^{(d)}$David Boyd, $^{(e)}$Robin Leadbeater, $^{(f)}$Pavol A. Dubovsky,  $^{(g)}$Forrest Sims, $^{(h)}$Daniel Dejean, $^{(i)}$\textit{West Challow Observatory} in England, $^{(j)}$\textit{Three Hills Observatory} in England, $^{(k)}$\textit{Vihorlat National Telescope} in Slovakia, $^{(l)}$\textit{Desert Celestial Observatory} in USA, and $^{(m)}$\textit{Labastide St Sernin} in France.
		\par}
\end{table*}

\section{Results and Discussion}\label{sec:results}

\subsection{Optical light curve}\label{lc}
\begin{figure}
	\centering
	\includegraphics[scale=0.62]{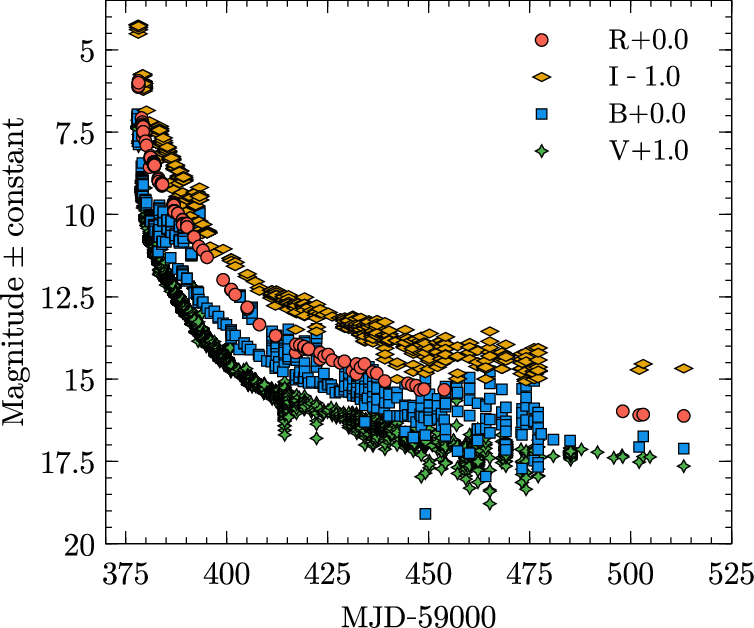}
	\caption{ \textit{BVRI} Light curves of Nova Her 2021 generated using optical data from AAVSO. Offsets are applied to the V and I bands (as indicated in the legend) for the sake of clarity.    
	}
	
	\label{fig:lightcurvenher21}
\end{figure}

The optical \textit{BVRI} light curve of nova V1674 Her over the initial 140 days, crafted using the AAVSO database, is shown in Fig.~\ref{fig:lightcurvenher21}. Analysing the light curve, we ascertain that the peak of brightness, $V = 6.13$, was attained on 2021 June 12.959 UT (JD 2459378.459), inline with \citep{2021ATel14704, 2021oodward}. 
As such, we adopted 2021 June 12.959 UT as the reference time for the outburst, denoted as $t_0$. Calculations for the characteristic times, $t_2$ and $t_3$, in V and B bands yield values of 0.904 and 1.935 days, and 1.191 and 6.387 days, respectively. It is notable that the B band has a more gradual rate of decline compared to the V band. This is because novae fade in brightness with age and become bluer in \textit{B-V} \citep{1987A&Avan_den_Bergh}. The $t_2 = 0.904$ days seems to be the quickest brightness decline time of any nova. This shows that the nova V1674 Her is the fastest classical nova ever observed, which is consistent with \citet{2021ATel14746, 2021oodward, 2022RNAASWoodward}. Other notable swift CNe include V838 Her, 1991 ($t_2 = 1$ day), V1500, 1975 ($t_2 = 2$ days), V1500 Cyg, 1975 ($t_2 = 2$ days), and V4160 Sgr, 1991 ($t_2 \sim 2$ days) \citep{2010AJStrope}. The optical light curve depicted in Fig.~\ref{fig:lightcurvenher21} demonstrates a smooth, rapid decline in brightness, suggesting that the nova did not produce an appreciable amount of dust.

\subsubsection{Reddening and distance calculation}  \label{rdc}
Using the established optical reddening law, which correlates interstellar extinction with colour excess as ($A_{\lambda}=R_{\lambda}E(B-V)$), we calculated the interstellar extinctions in the \textit{BVRI} -bands, corresponding to wavelengths of: $4400,~5500,~6540$, and $8060 ~\AA$, respectively. We adopted the total-to-selective extinction ratio ($R_{\lambda}$) values: $R_B =4.1$, $R_V=3.1$, $R_R=2.3$, and $R_I=1.5$ \citep{1994AJRichmond}. 
The reddening value for V1476 Her has been calculated as $E(B-V)=0.55$ \citep{2021ATel14704, 2021ATel14723Woodward}. Therefore, we employed the same value of $E(B-V) = 0.55$ in this study. Consequently, we derived the interstellar extinction values in the \textit{BVRI} -bands to be: $A_B =2.255$, $A_V=1.705$, $A_R=1.265$, and $A_I=0.825$. Subsequently, we computed the absolute magnitude ($M_{\lambda}$) employing the Maximum Magnitude-Rate of Decline (MMRD) equation given below \citep{1995ApJ-della-Valle}.
\begin{equation}
	M_{\lambda}=-7.92-0.81\arctan (\frac{1.32-\log t_2({\lambda})}{0.23}),
	\label{della}
\end{equation}
where $t_2({\lambda})$ corresponds to the $t_2$ values in the \textit{BVRI} -bands, derived from AAVSO photometric data. These values are calculated to be: $t_2(B)=1.149$ days, $t_2(V)=0.904$ days, $t_2(R)=1.96$ days, and $t_2(I)=1.226$ days. Using these values, we calculated the absolute magnitudes in each band to be: $M_{B}=-9.057$, $M_{V}=-9.046$, $M_{R}=-9.013$, and $M_{I}=-9.043$. Subsequently, we employed the distance modulus equation to calculate distance values, which is given by,
\begin{equation}
	m_{\lambda}-M_{\lambda}=5log_{10}\left(\frac{d_{\lambda}}{10} \right)+A_{\lambda},
	\label{eq:1}
\end{equation}
where $m_{\lambda}, d_{\lambda},$ and $A_{\lambda}$ are the apparent magnitude, distance, and extinction in the \textit{BVRI} -bands, respectively. The values of $m_{\lambda}$ at $t_2$ were derived from the AAVSO photometric data as follows: ~$m_{B}=8.962$ mag., ~$m_{V}=8.147$ mag., ~$m_{R}=7.901$ mag., and ~$m_{I}=7.248$ mag. Using these values, we calculated distance of  4.97 kpc in \textit{V} band, 5.60 kpc in \textit{B} band, 5.48 kpc in \textit{R} band, and 4.91 kpc in \textit{I} band. Combining all these, the average measured distance across the four bands is found to be $5.24\pm^{+0.36}_{-0.33}$ kpc. Our estimate is in a good agreement previous estimates, e.g. \citet{2021oodward} estimated the distance in the V band to be $4.75^{+1.3}_{-1.0}$ kpc from the combination of MMRD with intersection of the extinction versus distance curve, and \citet{2021AJBailer-Jones} estimated a distance of $6.0\pm^{+3.8}_{-2.8}$ kpc using \textit{Gaia}-EDR3.

\subsubsection{White dwarf mass and radius } \label{wdm}
The decline time $t_3$ can be utilized to estimate the mass of the WD, as at $t_3$, $\sim$85\% of the entire shell mass is expelled \citep{1981ApJShara}. This indicates that a lengthy $t_3$ signifies a substantial ejecta shell mass and a relatively small WD mass. To compute the WD mass ($M_{\text{WD}}$) of Nova V1674 Her, we employed the ensuing relationship, equation (\ref{t3}) (\citet{1992ApJLivio}):
	\begin{equation}\label{t3}
		 t_3=C\left[\dfrac{M_{\text{WD}}}{M_{ch}}\right]^{-1}\left[\left(\dfrac{M_{\text{WD}}}{M_{ch}}\right)^{-2/3}-\left(\dfrac{M_{\text{WD}}}{M_{ch}}\right)^{2/3}\right]^{3/2} \text{days},
	\end{equation}
where $C = 51.3$ (a constant calibrated from V1500 Cyg) \citep[][and references therein]{1992ApJLivio}, $M_{ch}=1.4~M_{\sun}$ (Chandrasekhar mass limit), and $t_3=1.935$ days (Section \ref{lc}). By solving equation (\ref{t3}) and plugging in the relevant values, we arrive at an estimate of $M_{\text{WD}}\sim1.36M_{\sun}$. Such a substantial mass, nearing the Chandrasekhar limit, is necessary to account for the high ejection velocities ($\sim$ 6000 \kms) \citep{2021oodward}. \citet{2021ApJDrake} also estimated $M_{\text{WD}}>1.05M_{\sun}$. The mass of WDs are a principal distinction between ONe and CO type novae. In case of ONe novae  $M_{\text{WD}}>1.2~M_{\sun}$, whereas in case of CO novae  $M_{\text{WD}} < 1.2M_{\sun}$ \citep{1998PASPGehrz}. A more recent work by \citet{2015MNRASDoherty, 2000A&AWeidemann} reported that the WD mass in ONe novae is $>1.0M_{\sun}$.  Using the following expression \citep{warner1995}, 
	\begin{equation}
		\label{rad}
		R_{\text{WD}} \approx 0.90\left[1-\dfrac{M_{\text{WD}}}{M_{ch}}\right]^{1/2}~R_{\earth}, 
	\end{equation}
	the possible smallest radius of the WD $R_{\text{WD}}$ is $\sim 969$ Km ($\sim 0.15~R_{\earth}$), where $R_{\earth}$ denotes the radius of the Earth.

\subsection{Spectral evolution }\label{sec3}
\begin{figure*}
	\centering
\hspace*{0.2cm}	\includegraphics[scale=0.63]{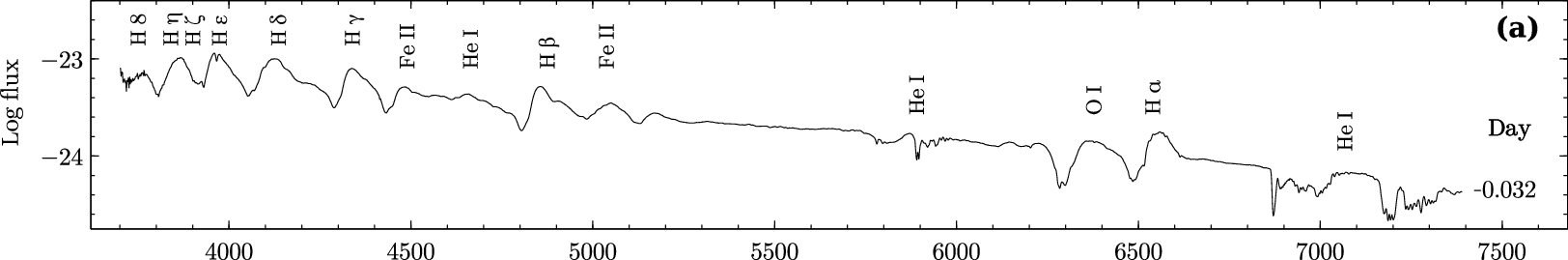}\\
\vspace{0.5cm}
\includegraphics[scale=0.635]{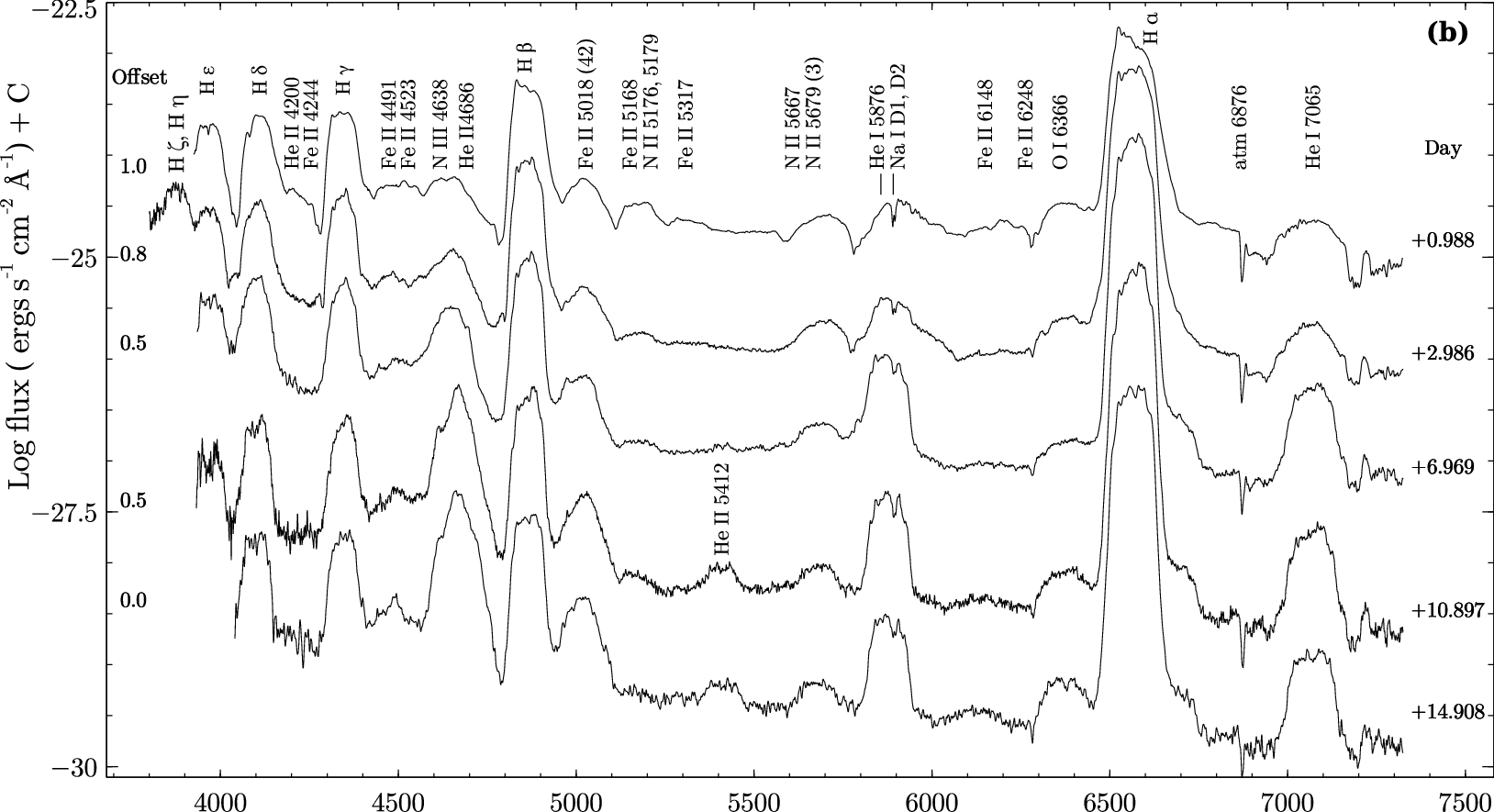}\\
 \vspace{0.5cm}
\includegraphics[scale=0.635]{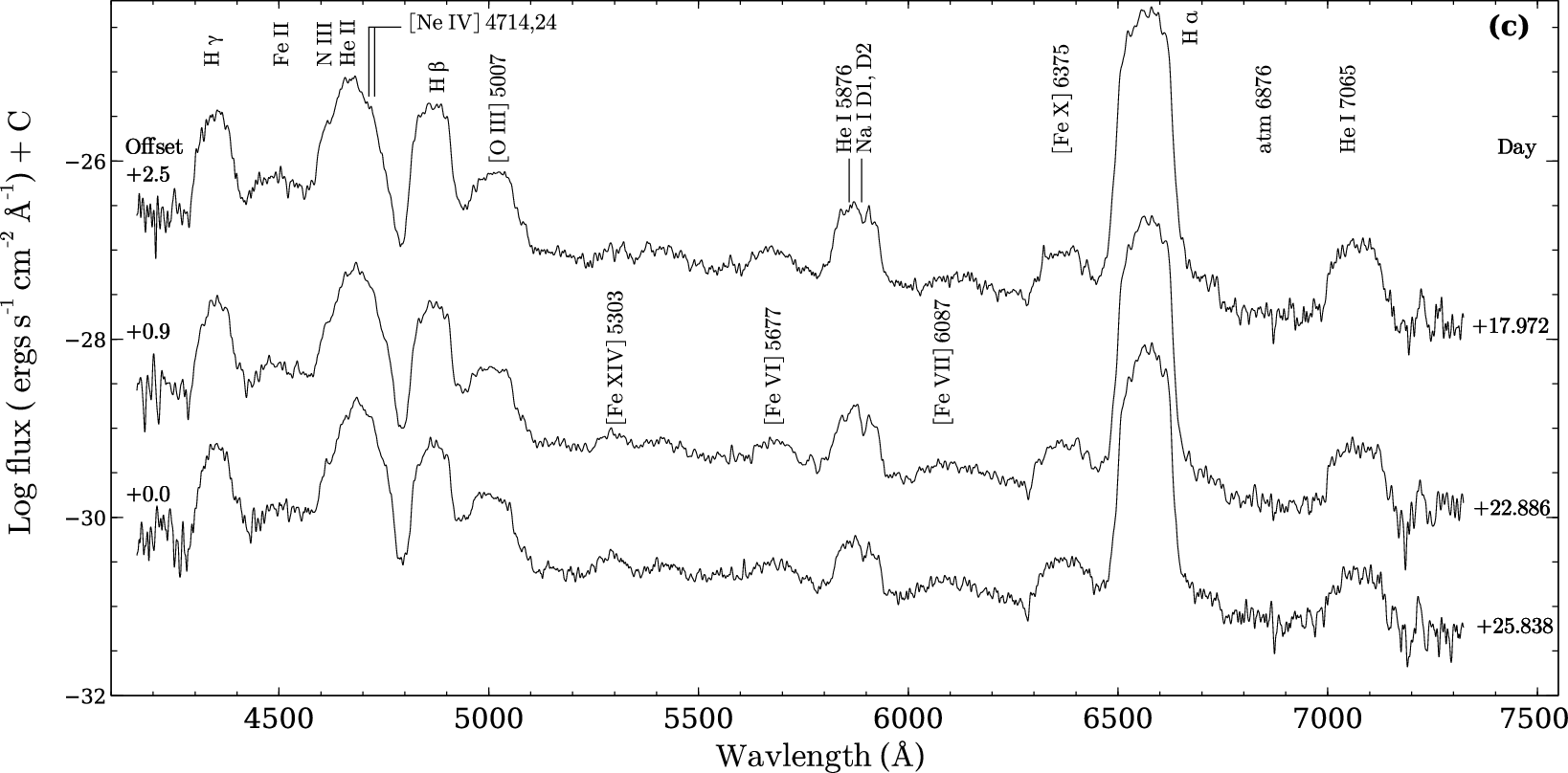}
\caption{ Evolution of V1674 Her's medium-resolution optical spectra acquired: (a) on 2021 June 12.923 UT (-0.032 days); (b) from 2021 June 13.947 (+0.99 days) to June 27.867 UT (+14.91 days); and (c) from 2021 June 30.931 (+17.97 days) to July 07.866 UT (+25.84 days). For the sake of clarity, the spectra in (b) and (c), are offset by the amount indicated on the left side of the figure. The ordinate scale is logarithmic to emphasise the visibility of weaker features relative to the emission feature of \ion{H}{$\alpha$}. \lq{atm}\rq~ stands for atmospheric absorption observed at around 6876 $\AA$.}
	\label{fig:MRS}
\end{figure*}

\begin{figure*}
	\centering
	\includegraphics[scale=0.63]{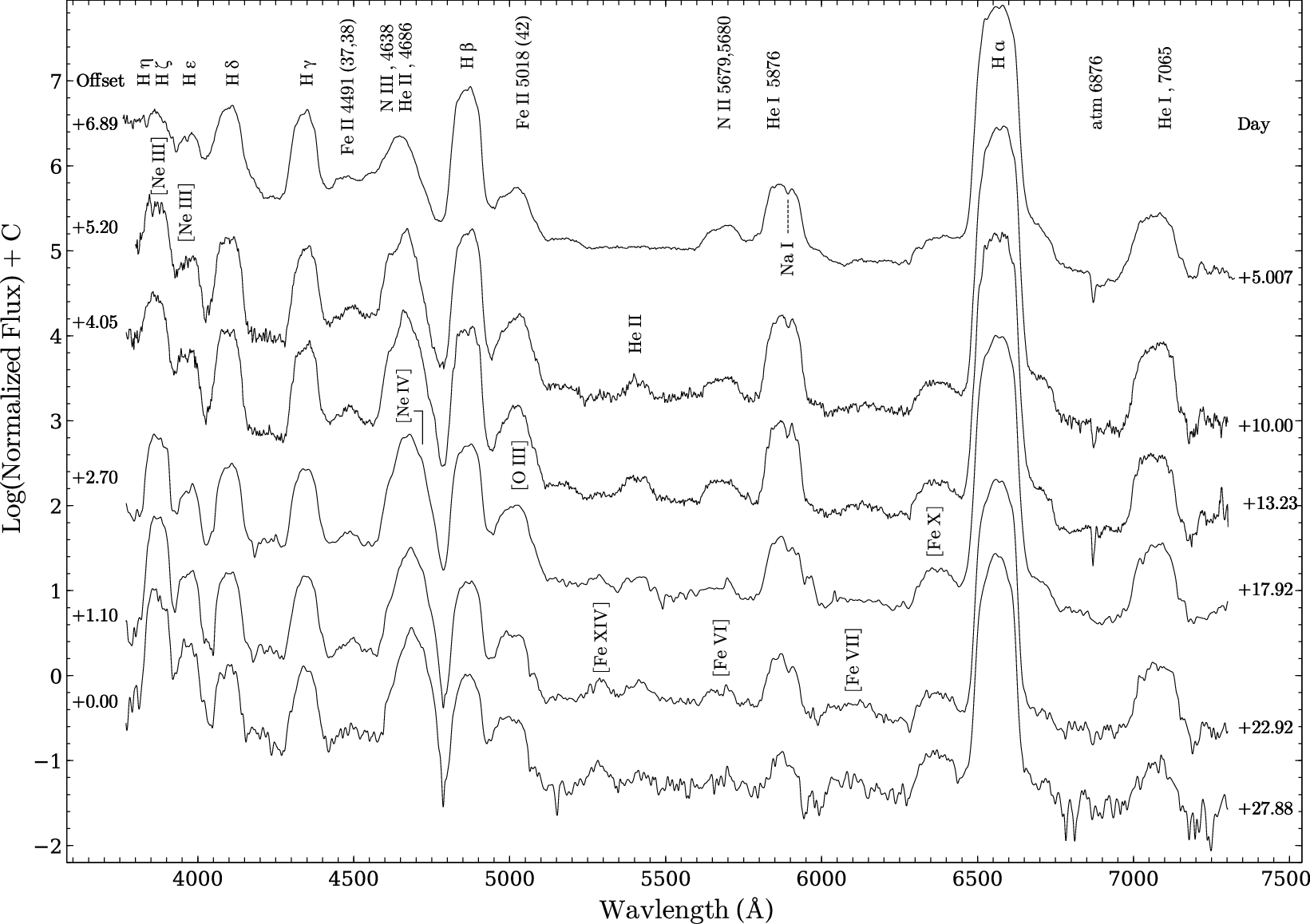}
	\caption{Evolution of V1674 Her's low-resolution optical spectra  obtained from 2021 June 17.985 (5.007 d) to July 09.894 (27.880 d). For the sake of clarity, the spectra are offset by the amount indicated on the left side of the figure. The ordinate scale is logarithmic to emphasize the visibility of weaker features relative to the emission feature of \ion{H}{$\alpha$}. 'atm' stands for atmospheric absorption.
	}
	\label{fig:LRS}
\end{figure*}

The spectral evolution of Nova V1674 Her during the initial month, spanning wavelengths of: $\sim 3700 - 7400$ {\AA}, $3920 - 7330$ {\AA}, $4170 -7330$ {\AA} and $3800 - 7300$ {\AA}, is presented in Fig.~\ref{fig:MRS} (a), ~\ref{fig:MRS} (b), ~\ref{fig:MRS} (c) and \ref{fig:LRS}, respectively. All the notable emission features are identified. Fig.~\ref{fig:Pcygni_profile} illustrates the blue shifting of the P Cygni profile over time. All spectra are normalised to the \ion{H}{$\beta$} line and corrected for reddening using $E(B-V) = 0.55$. 
The flux and widths of emission lines profiles were measured interactively using the tools provided by \textsc{IRAF} (Image Reduction and Analysis Facility)\footnote{\url{http://iraf.noao.edu/}}. Analysis of dereddened spectra over the initial twenty-eight days, showed a notable reduction in observed flux during this period. Within the initial month of the outburst, the spectral evolution of nova V1674 Her underwent four distinct phases: pre-maximum, early decline, nebular, and coronal stages.

\subsubsection{Pre-maximum Stage}\label{final}
The pre-maximum spectrum, observed on 2021 June 12.927 UT (i.e., 0.032 days before the visual maximum), is illustrated in Fig.~\ref{fig:MRS} (a). This spectrum prominently featured Balmer, \ion{Fe}{ii} and \ion{He}{I}  lines, accompanied by distinct narrow and strong P Cygni profiles. The spectrum exhibited relatively weaker hydrogen Balmer emission lines, encompassing \ion{H}{8}, 3771 {\AA}. This was because at this stage, a significant portion of the line-forming region was optically thick, leading to the attenuation of emission lines. The development of a large, optically thin region before reaching visual maximum is an unusual occurrence for a nova \citep{2001MNRAS-Schwarz}. We noticed that the \ion{H}{$\eta$} 3835 \AA~ and \ion{H}{$\zeta$} 3889 \AA~ lines appeared blended at this stage.  The P-Cyg profiles of \ion{H}{$\alpha$}, \ion{He }{i}, \ion{H}{$\beta$}, and \ion{H}{$\gamma$} were blue-shifted from the central maxima by $-3598.88 \pm 42$, $-3388.81 \pm 46$ , $-3519.73 \pm 50$, and $-3575 \pm 60$ \kms, respectively. Notably, these blue shifts were relatively lower for each line, as the blue shift of the P Cygni absorption line tends to increase after the outburst (refer to Fig.~\ref{fig:Pcygni_profile}).

\begin{figure}
	\centering
	\includegraphics[scale=0.7]{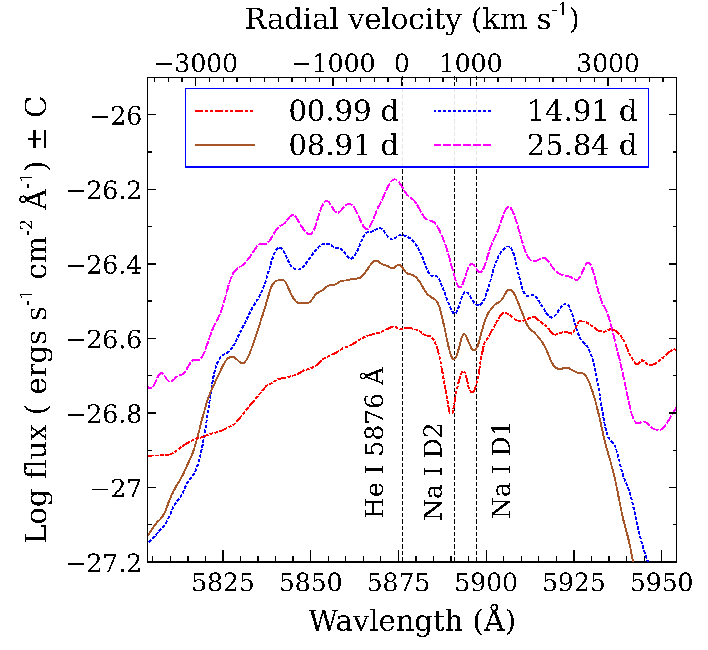}
	\caption{Emission line profiles of \ion{He}{i} 5876 \AA~ on 2021 June 13 (0.99 d), June 21 (8.91d), June 27 (14.91 d), and July 07 (25.84 d). These emission lines show the \ion{Na}{i} D1 and D2 doublets clearly. The Gray vertical dashed line represents radial velocity, $V_{rad}=0$ \kms (i.e., 3876 \AA), whereas the black dotted lines represent \ion{Na}{I} \textsc{D2} and \textsc{D1} at $V_{rad}=773$ and $1078$ \kms, respectively, in left-to-right order. }
	\label{fig:NaD12}
\end{figure}

In addition to the P Cygni absorption features, a few other notable absorption features have also been observed among the considered spectra, including the pre-maximum spectrum. Firstly, the spectra displayed two adjacent absorption features on top of the \ion{He}{i} 5876 \AA~ line, namely: \ion{Na}{I} \textsc{D1} and \ion{Na}{I} \textsc{D2} lines. Notably, the \ion{Na}{I} \textsc{D2} line consistently appeared stronger than the \ion{Na}{I} \textsc{D1} line in all cases (see Fig. \ref{fig:NaD12}). These absorption lines might be attributed to the in-homogeneous nature of the ejecta, consisting of discrete emitting and absorbing clouds with a low filling factor \citep{1991williams}. Despite a slight variation in the strength of these absorption lines, the feature remains consistently present in every spectrum acquired within the initial month following the outburst. These features were evident in our medium-resolution spectra, as shown in Fig. \ref{fig:MRS}, \ref{fig:LRS}, \ref{fig:NaD12}, and \ref{fig:line_profile}.	The \ion{Na}{I} \textsc{D1} and \textsc{D2} absorption lines were red-shifted with respect to the rest frame (radial velocity, $V_{\text{rad}}=0$\kms) of the \ion{He}{I} 5876 \AA~ line, by $+1078.00 \pm ^{+49.31}_{-64.09}$ and $+773.00 \pm ^{+57.72}_{-34.80}$ \kms, respectively. These values were obtained by averaging the radial velocities of corresponding \ion{Na}{I} \textsc{D1} and \textsc{D2} lines observed on days: 0.99, 8.91, 14.91, and 25.84. Secondly, all the spectra, including the pre-maxima stage spectrum, consistently exhibit a telluric feature at a wavelength of $6876~\AA$ (see Fig.~\ref{fig:MRS}). Over time, the intensity of these absorption features diminished; nevertheless, their distinctive presence persisted in all spectra.

\subsubsection{Early Decline Stage (0 to $\sim$10 days)}
The set of spectra presented in Fig.~\ref{fig:MRS} (b) are taken during the early decline phase.
During this phase, the spectra displayed prominent and broad recombination lines, encompassing Balmer and Helium lines, as well as various iron multiplets. These spectra undeniably exhibited the typical features of \ion{Fe}{ii} novae, characterised by numerous low-excitation \ion{Fe}{ii} lines \citep{2012AJWilliams}. Transitioning from the pre-maximum stage to the early decline stage, these recombination lines, notably the Balmer lines, became considerably stronger in relation to the continuum level. However, as time progressed, these emission lines gradually reduced in width, a trend commonly observed in novae initiated by thermonuclear runaways (TNRs) \citep{1996ApJshore, 2004A&ACassatella, 2012AJ-williams}. This phenomenon is attributed to the diminishing contribution of the outermost ejecta region as they move away from the central object \citep{2004A&ACassatella}. As the expanding ejecta thin out, the ionization of the ejecta intensifies, owing to the ongoing illumination and ionization from the hot central ionizing source.

\begin{figure}
	\centering
	\includegraphics[scale=0.54]{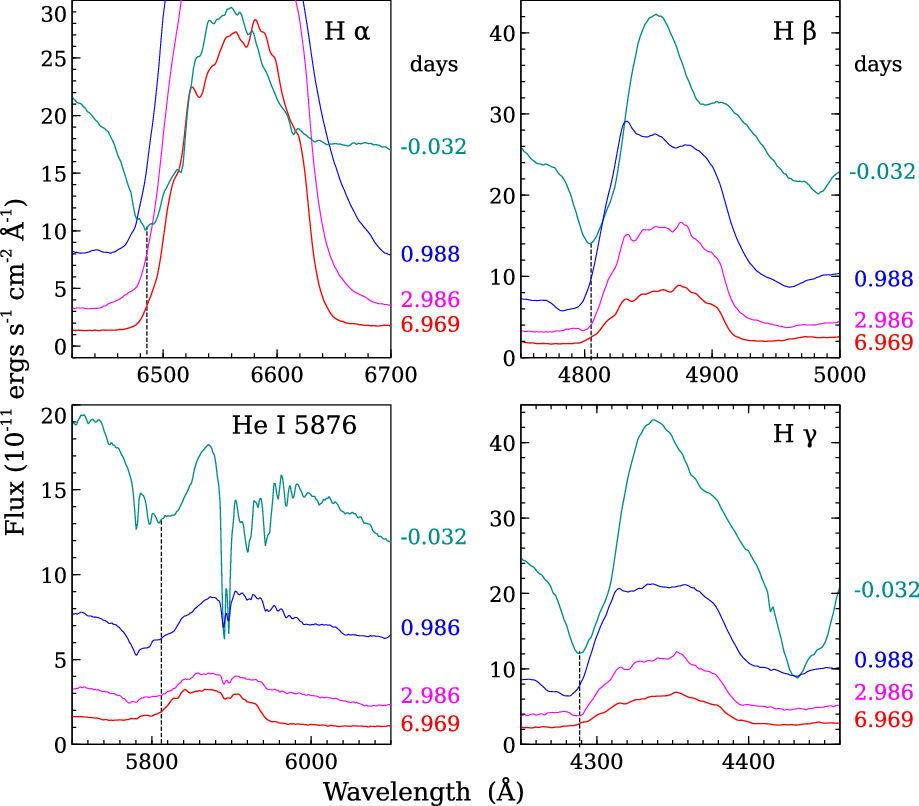} 
	\caption{ Emission line profiles of \ion{H}{$\alpha$}, \ion{He }{I}, \ion{H}{$\beta$}, and \ion{H}{$\gamma$}, along with their associated P Cygni profiles, displayed a blue shift as time progressed. The spectra are flux calibrated and deredenned with an extinction value \textit{E(B-V)} of 0.55. In order to combine everything, we subtracted constants 19, 30, 35, and 50 from the flux component of the day -0.032 emission line profiles of the \ion{H}{$\alpha$}, \ion{He }{I}, \ion{H}{$\beta$}, and \ion{H}{$\gamma$} lines, respectively. All these constants are in the same unit of flux. 
	}
	\label{fig:Pcygni_profile}
\end{figure}

Fig. \ref{fig:Pcygni_profile} illustrates the broad emission lines of \ion{H}{$\alpha$}, \ion{He }{I} 5876, \ion{H}{$\beta$}, and \ion{H}{$\gamma$}, accompanied by their respective P Cygni profiles on days -0.032, +0.99, +2.99, and +6.969. On day 0.988, these P Cygni absorption lines exhibited blue shifts of $-4988.93 \pm 42$, $-4849.57 \pm 46$, $-4841.05\pm 39$, and $-3998.23 \pm62$ \kms, relative to the centre of emission line of  \ion{H}{$\alpha$}, \ion{He}{I}, \ion{H}{$\beta$}, and \ion{H}{$\gamma$}, respectively. Notably, the P Cygni profiles exhibited greater shifts compared to the pre-maximum stage. This shift is attributed to the increasing ejection velocity of novae with distance at least during the free expansion phase. Gradually, the P Cygni absorption features diminished and vanished in later days. 

\begin{figure}
	\centering
	\includegraphics[scale=0.7]{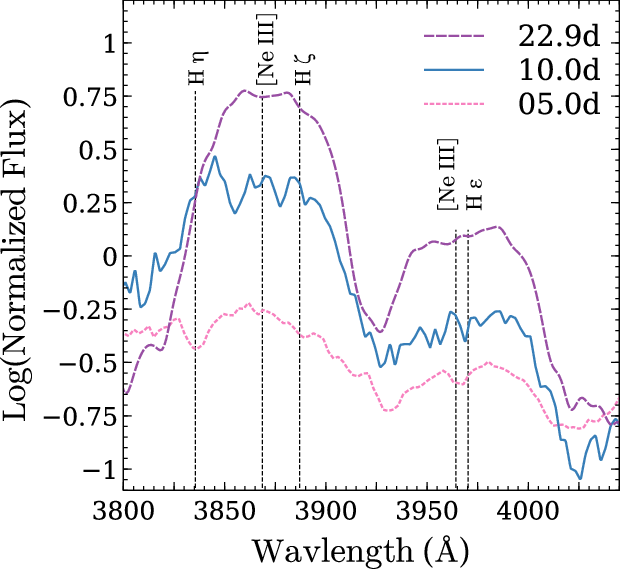}
	\caption{Emission features of Balmer lines (\ion{H}{$\eta$} 3835 \AA, \ion{H}{$\zeta$} 3889 \AA, and \ion{H}{$\epsilon$} 3970 \AA), on 2021 June 17.99 (5.0 d), June 22.99 (10.0 d), and July 04.925 (22.92 d). 
 Starting from day 10.0 and onward, \ion{H}{$\eta$} and \ion{H}{$\zeta$} became blended with [\ion{Ne}{III}] 3869.07 \AA, while \ion{H}{$\epsilon$} became blended with [\ion{Ne}{III}] 3967.69 \AA. }
	\label{fig:NeIII15}
\end{figure}

\begin{figure}
	\centering
	\includegraphics[scale=0.7]{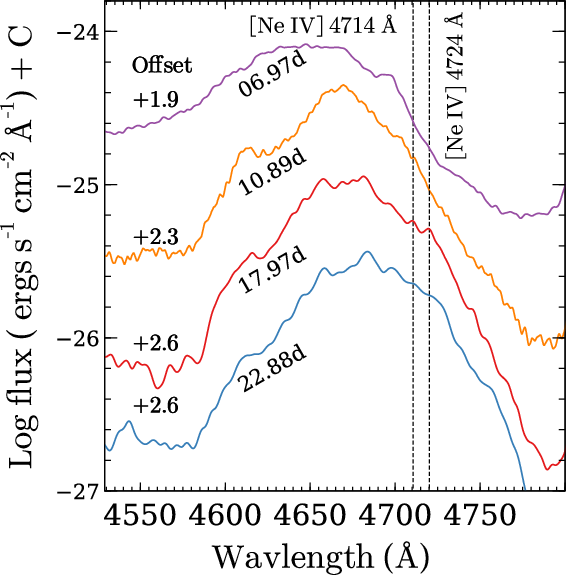}
	\caption{Emission line profiles of \ion{N}{iii} 4638 \AA,  and \ion{He}{iii} 4686 \AA on 2021 June 19.93 (06.97 d), June 23.86 (10.99d), June 30.93 (17.97 d), and July 04.91 (22.88 d). The black dashed vertical lines indicate the emergence of [\ion{Ne}{IV}] 4714 and 4724 \AA~ lines starting from at least day 17.97 and onward. }
	\label{fig:NeIV13}
\end{figure}

\subsubsection{Nebular Phase ($\sim$ 10 to 22 days)}\label{nebu}
In this phase, a rapid reduction in the intensity of neutral emission lines has been observed. For instance, the measured flux in the \ion{H}{$\alpha$}, \ion{H}{$\beta$}, \ion{H}{$\gamma$}, \ion{H}{$\delta$}, \ion{He}{i} 5876 \AA, and \ion{He}{i} 7065 \AA~ emission lines is 3.69, 3.92, 4.00, 4.18, 3.8, and 3.22 times lower on day 10.897 than it was on day 6.969, respectively. This swift spectral progression indicates that almost all of the ejecta's clumps, have undergone ionization, signifying that the ejecta has entered the optically thin phase, commonly known as the \enquote{nebular stage}. During this phase, the collisionally excited nebular lines become prominent in the expanding ejecta's spectra \citep{2001MNRAS-Schwarz}. On day 10.00, the Balmer lines, \ion{H}{$\eta$} and \ion{H}{$\zeta$}, became prominent in the observed spectra (see Fig.~\ref{fig:LRS}). This is due to the emergence of forbidden emission lines, such as [\ion{Ne}{iii}] 3869 \AA, in the spectrum (see section \ref{ec}, for the detail). Notably, the blended line at the spot of \ion{H}{$\eta$}, \ion{H}{$\zeta$}, and [\ion{Ne}{iii}] 3869 \AA~ stands out as the strongest emission line on this date, except for \ion{H}{$\alpha$}. These spectral characteristics often indicate the onset of the nebular phase. Similarly, \ion{H}{$\epsilon$} showed a significant increase in intensity, due to the appearance of [\ion{Ne}{iii}] 3968 \AA (see section \ref{ec}, for the detail). In Fig. \ref{fig:NeIII15}, we provide a visualisation of how the intensity and structure of the emission lines changed between day 5.00, 10.00, and 22.90. These two lines, [\ion{Ne}{iii}] at 3869 and 3968 \AA, were previously identified in nova V1674 Her on day 17.0 \citep{2021ATel14746, 2022RNAASWoodward}.

The \ion{N}{iii} 4638 \AA~ and \ion{He}{ii} 4686 \AA~ emission lines displayed notably low intensity relative to the continuum level in the initial stages. However, they later exhibited a significant increase, and on day 14.908, they became the second strongest emission line after \ion{H}{$\alpha$}.  This rapid evolution of the profile wasn't only due to the contributions of \ion{N}{iii} 4638 \AA, and \ion{He}{ii} 4686 \AA, but also the [\ion{Ne}{iv}] 4714, 4724 \AA~ lines on the right side of the profile (see the upper spectra in Fig.~\ref{fig:MRS} (c) and ~\ref{fig:NeIV13}). Although the [\ion{Ne}{iv}] 4714, 4724 \AA~ lines might have appeared earlier, it was on 2021 June 30 (17.972 d) that both forbidden neon lines became clearly observable. The presence of these neon lines in the spectra has also been reported by various researchers; for instance,  \citet{2021ATel14746} reported the detection of [\ion{Ne}{iv}] 4714 and 4724 \AA~ on day 15. Similarly, on day 17.92 (2021 June 30.90) we observed a notable increase in intensity in the line profile identified as \ion{Fe}{ii} 5018 \AA~ during the early decline phase. This increase is mainly due to the emergence of [\ion{O}{iii}] 5007 \AA~ in the spectrum (see Section \ref{ec} for details). The appearance of the [\ion{O}{iii}] line in the spectrum is usually considered a confirmation of the onset of the Nebular phase in the system. The presence of [\ion{O}{iii}] 5007 \AA~ was also observed by \citet{2021RNAASRudy} in the spectrum of day 15.00 (2021 June 28).

\subsubsection{Coronal Phase (after $\sim$ 22 days)} \label{coro}
The bottom two spectra in Fig.~\ref{fig:MRS} (c) and \ref{fig:LRS} illustrate the emergence of several coronal emission lines commencing around day 22.89. It is possible that certain lines might have emerged prior to 22.89. For example, a broad spectral line was observed on 2021 June 30, at $\sim$ 6375 \AA, which aligns with the typical location of the coronal line [\ion{Fe}{x}]. However, uncertainty remained whether this line is primarily influenced by the oxygen line, which appeared in the initial spectra phases, or by the forbidden iron lines resulting from an increase in ejecta temperature. From day 22.89 onwards, well-resolved higher forbidden ionization lines of iron, such as [\ion{Fe}{xiv}] 5303, [\ion{Fe}{vi}] 5677 \AA, [\ion{Fe}{vii}] 6087, and [\ion{Fe}{x}] 6375 \AA, became prominent in the spectra. According to the Tololo spectral classification outlined in \citet{1991williams}, it is explained that under any other circumstance of line strength, the nova spectrum can be considered in the coronal phase once the emission line [\ion{Fe}{x}] 6375 \AA~ has appeared and is stronger than [\ion{Fe}{vii}] 6087 \AA. These conditions are met for the case of V1674 Her on day 22.886 and thereafter. Therefore, based on spectral features and the appearance of forbidden emission lines of high ionization, we conclude that Nova Her 2021 has already entered the coronal phase by day 22.

According to \citet{2021oodward}, the nova V1674 Her entered the coronal phase 11.5 days after the outburst (on 2021 June 24) in the IR band, marking the earliest coronal phase onset of any recorded classical novae. Prior to the discovery of V1674 Her, two of the fastest novae known were V1500 Cyg (with $t_2=2$ days) and V838 Her (with $t_2=2$ days), which entered the coronal phase on days 29 and 17 following the outburst, respectively \citep{1990AJBenjamin, 1993BASI_Chandrasekhar}. Hence, it is unsurprising that Nova V1674 Her, 2021, exhibits strong coronal emission lines at a quicker rate than slower novae, some of which barely generate these lines \citep{1978ApJFerland, 1990AJBenjamin}. However, optical spectra acquired during this period did not confirm the rapid onset of the coronal phase. This divergence is attributed to the lower opacity of IR compared to optical and UV, resulting in the appearance of coronal lines in IR earlier than in the optical range \citep{1998PASPGehrz}.

\subsubsection{Line Profiles}\label{lp}
\begin{figure*}
	\centering
	\includegraphics[scale=0.74]{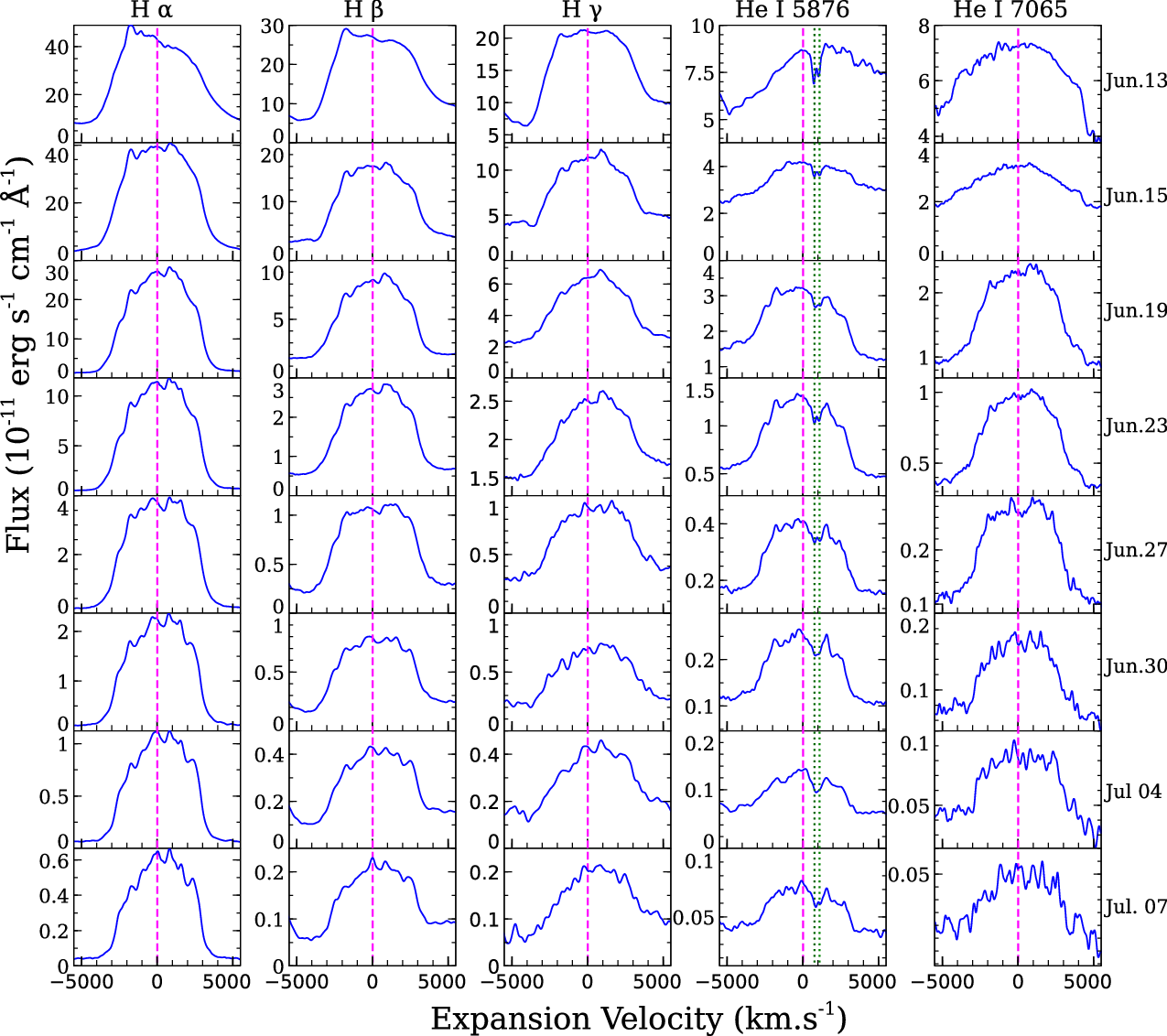}
	\caption{ Emission line profiles for \ion{H}{$\alpha$}, \ion{H}{$\beta$}, \ion{H}{$\gamma$}, \ion{He}{I} 5876 \AA, and \ion{He}{I} 7065 \AA~ observed on 2021 June 13.95, June 15.95, June 19.93, June 23.86, June 27.87, June 30.93, July 04.91, and July 07.87 UT.  Interstellar extinctions are corrected with \textit{E(B-V)} = 0.55. The horizontal axis represents expansion velocity in \kms, while the vertical axis represents observed flux in \ergscmA~ units. The vertical broken magenta line denotes the centre (rest frame) of each profile. The green dotted vertical lines in column four indicate \ion{Na}{I} D1 (right) and D2 (left) lines.}
	\label{fig:line_profile}
\end{figure*}

In Fig.~\ref{fig:line_profile}, we illustrate the evolution of recombination line profiles for the \ion{H}{$\alpha$}, \ion{H}{$\beta$}, \ion{H}{$\gamma$}, \ion{He}{I} 5876 \AA, and \ion{He}{I} 7065 \AA~ lines. After the outburst, these lines exhibited intricate profiles characterised by a broad central component accompanied by multiple sub-peaks on the top of each line profiles. Broad lines denote fast-moving shell ejecta \citep{2022ApJNaito}, while broad wings and peaks in emission lines indicate density and velocity in-homogeneities (i.e., slow-moving clumps embedded in diffuse gas) \citep{2007ApJSchwarz}.
On day +0.99, the emission line profiles of \ion{H}{$\alpha$} and \ion{H}{$\beta$} appeared stronger on the blue side than the red side of the profile, resulting in an inclined top of the line profiles (see Fig.~\ref{fig:line_profile}). This indicates asymmetric dispersion of the ejecta along the line of sight, involving both approaching and receding components. However, this condition didn't last long; by day +2.96, the red side of the \ion{H}{$\alpha$} and \ion{H}{$\beta$} line profiles appeared somewhat stronger than the blue side. Further details are available in the morpho-kinematic section of this paper (Section \ref{mkm}). 

\begin{figure}
	\centering
	\includegraphics[scale=0.58]{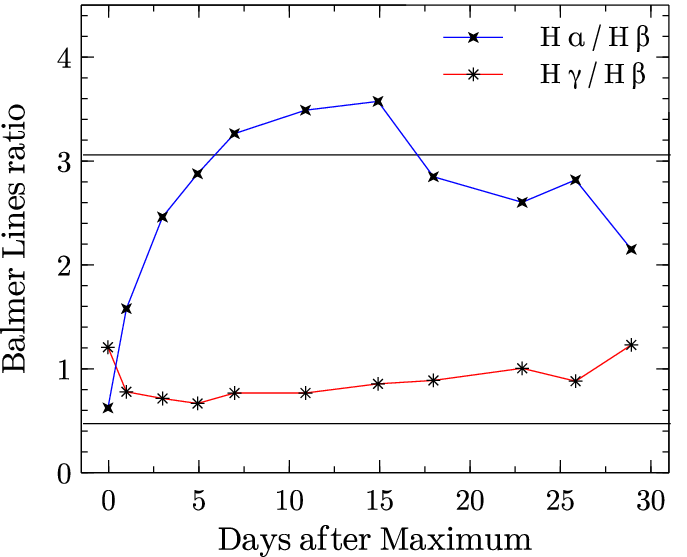}
	\caption{ Evolution of ratios of reddening corrected Balmer line fluxes (\ion{H}{$\alpha$} / \ion{H}{$\beta$} and \ion{H}{$\gamma$} / \ion{H}{$\beta$}), from day +0.988 to 28.924. The horizontal lines represent the Balmer lines ratio for case B at $10^4$ K. The upper and lower horizontal lines represent the theoretical estimation of \ion{H}{$\alpha$} / \ion{H}{$\beta$} and \ion{H}{$\gamma$} / \ion{H}{$\beta$}, respectively. }
	\label{fig:Balmer_ratio}
\end{figure}

The evolution of Balmer line flux ratios of \ion{H}{$\alpha$} and \ion{H}{$\gamma$} relative to \ion{H}{$\beta$} from day +0.988 to +28.924 is shown in Fig.~\ref{fig:Balmer_ratio}.  In the graph, we have included two horizontal lines to comprehend how closely the observational results align with the theoretically estimated value of the Balmer line flux ratio for Case B at $10^4$ K, as calculated by \citep{1989agnaOsterbrock}. The upper and lower horizontal lines represent \ion{H}{$\alpha$}/\ion{H}{$\beta$} and \ion{H}{$\gamma$}/\ion{H}{$\beta$} ratios, having values of 3.08 and 0.46 respectively \citep{1989agnaOsterbrock,2001MNRAS-Schwarz}. As depicted in the Fig.~\ref{fig:Balmer_ratio}, the \ion{H}{$\alpha$}/\ion{H}{$\beta$} ratio during the period between days +6.969 and +14.908, as well as the overall \ion{H}{$\gamma$}/\ion{H}{$\beta$} ratio, exceeds the expected value based on recombination under optically thin circumstances. This stipulates the presence of a high-density region contributing to the emission. This result is inline with the photoionization model result discussed in Section \ref{ed}. 

\subsubsection{V1674 Her as a neon nova} \label{nn}
V1674 Her exhibited a variety of distinct characteristics of neon novae. For instance, Firstly, neon novae are distinguished by a substantial WD mass. The estimated WD mass of V1674 Her is $~\sim~1.36~M_{\sun}$ (Section \ref{wdm}), aligning with the typical WD mass range for neon novae. Secondly, ONe novae tend to cluster more closely towards the galactic plane than lower-mass CO novae \citep{1992A&A}. Similarly, V1674 Her is positioned at galactic coordinates ($l=048^{\circ}.707$, $b= +06^{\circ}.3114$), in closer proximity to the galactic plane, as anticipated. Thirdly, the optical light curve in Fig.~\ref{fig:lightcurvenher21} reveals a smooth and rapid decline in brightness without any evident signs of dust formation. \citet{2021ATel14741woodward, 2021ATel14728Woodward} also noted the absence of significant dust formation in the first two months following the nova's outburst. According to \citet{2013ApJWilliams}, fast neon novae tend to produce minimal dust, while CO novae generate more dust due to their lower-mass WDs, resulting in higher mass ejections at slower speeds \citep{1998PASPGehrz}. Fourthly, neon novae exhibit prominent neon emission lines, and their ejecta often achieve higher velocities compared to CO novae \citep{1986ApJStarrfield, 1996vanlandingham, 2016ApJHachisu}. As clearly depicted in Fig.~\ref{fig:LRS}, neon lines began to emerge in the spectra along with Balmer and helium lines around day +10.00 following the outburst. The \ion{Fe}{ii} multiplets gradually diminish, while neon lines become stronger than most of the permitted lines, except \ion{H}{$\alpha$}. This transition suggests that the nova shifted from an \ion{Fe}{ii}-dominant type to a neon nova on day 11, consistent with previous studies \citep{2021ATel14746, 2021oodward, 2022RNAASWoodward}. The estimated elemental abundances obtained from photoionization modelling also support the gradual decrease in iron content and simultaneous increase in neon abundance (Section \ref{ec}). Such hybrid characteristics are atypical within the Nova system. 

\subsection{Photoionization Model}\label{pma}
In this study the photoionization code \textsc{cloudy}\footnote{\url{https://trac.nublado.org/}} (V17.02; \citet{ferland17}) was employed to simulate the emission line spectra of the nova V1674 Her. Previous studies, including \citet{2002ApJSchwarz, 2003AJShore, 2005ApJVanlandingham,  2007ApJSchwarz} have also used \textsc{cloudy} to study various kinds of novae. This code models the physical conditions of non-equilibrium gas clouds exposed to external radiation fields, predicting emission lines spectrum based on assumptions about the gas's physical condition (ionization, density, temperature, and chemical composition). Through a self-consistent algorithm, \textsc{cloudy} simultaneously solves thermal, statistical, and chemical equilibrium equations for input parameters, estimating both intensities and column densities of numerous spectral lines across the electromagnetic spectrum. Major ionization and recombination processes are accounted for in \textsc{cloudy}, such as photoionization, Auger ionization, collisional ionization, charge transfer, and three-body recombination. Model-generated spectra are compared to observations to deduce physical and chemical conditions. Input parameters comprise temperature (T), luminosity (L), hydrogen number density ($n$), filling factor, covering factor, elemental abundances, and inner and outer radii of the surrounding ejecta. For synthetic model spectra, we included these input parameters and the abundances of elements with observed emission lines, whereas other elements were maintained at their solar values according to \citep{2010ApGrevesse}.

We initiated our modelling by assuming a central ionizing source with a high temperature ($T > 10^4$ K), whose physical properties are determined by the temperature and luminosity parameters across all of our models. This central ionizing source is enveloped by a spherically symmetric ejecta, whose density is governed by the parameter of hydrogen number density. The density distribution of this ejecta follows a power-law profile dependent on radius, as proposed by \citet{1976MNRASBath}, i.e., $\rho \propto r^{\alpha}$, where $\alpha$ represents the power-law index. The degree of clumpiness, indicating the proportion of gas within the total volume, is established by the filling factor parameter, denoted as $\textrm{f(r)}$. Notably, \textsc{cloudy} employs radial-dependent hydrogen density and filling factor, which are defined as follows:
\begin{equation}
	\frac{n(r)}{n(r_0)}=\bigg(\frac{r}{r_0}\bigg)^{\alpha}~\text{cm}^{-3},
\end{equation}
\begin{equation}
	\frac{f(r)}{f(r_0)}=\bigg(\frac{r}{r_0}\bigg)^{\beta},
\end{equation}
where $\alpha$ and $\beta$ represent the exponents of the power laws, $r_0$ denotes the inner radius, and $\textrm{n(r)}$ and $\textrm{f(r)}$ indicate the hydrogen density and filling factor at a radius $r$, respectively.
The shell's density is governed by a hydrogen density parameter with a power-law density profile and an exponent of -3, chosen to maintain a steady mass per unit volume throughout the model shell ($\dot{\textrm{M}}=$ const.) and a linear velocity law ($\textrm{v} \propto \textrm{r}$). \citet{2006A&AEderoclite, shore_2008} estimated that the maximum filling factor of a nova shell ejecta, which typically occurred at the beginning of the shell expansion, was 0.1.  This value could decrease as the shell expands. Thus, as our model covered the first month following the outburst, we adopted a filling factor value of 0.1, resulting in a power law exponent of $\beta = 0.0$. The inner and outer shell radii were determined using the minimum and maximum expansion velocities obtained from the \textsc{FWHM} of all emission lines and the time between the outburst and discovery. The minimum velocity for each of the six selected epochs ranged between 2893.5 and 3290 \kms, and the maximum velocity ranged between 5260 and 6566.6 \kms.

The set of synthetic spectra were generated by simultaneously varying all the aforementioned input parameters in smaller increments across a wide range of values. Temperature was varied between $10^{4.5}$ and $10^6$ K, luminosity between $10^{36}$ and $10^{41}$ \ergs, and ejecta density between $10^{6.5}$ and $10^{13} , \text{cm}^{-3}$, along with elemental abundances. Multiple test models were iterated across all epochs numerous times before arriving at the final model. Initially, the models were visually assessed, and those not aligning with the observed spectrum were discarded. Ultimately, to assess the fitting quality, we computed the $\chi^2$ and $\chi^2_{\text{red}}$ values of the model, expressed by the following relation:
\begin{equation}
	\chi^2=\sum_{i=1}^{n}\frac{(M_i-O_i)^2}{\sigma^2}, \qquad \text{and}\qquad \chi_{\text{red}}^2=\frac{\chi^2}{\nu},
\end{equation}
where $O_i$ and $M_i$ represent the ratios of observed and modelled line fluxes to the \ion{H}{$\beta$} line flux, respectively, and $\sigma_i$ denotes the error in the observed flux ratio. The degrees of freedom, $\nu$, are given by $n - n_p$, where $n$ is the count of observed lines and $n_p$ is the number of free parameters. Typically, $\sigma$ falls within the range of 10 to 30 percent, depending on its strength relative to the continuum and its potential for blending with other spectral lines \citep{Helton10}. Interactive flux measurements were conducted by fitting Gaussians using the \textit{splot} task within the \textit{onedspec} package in \textsc{IRAF}. An ideal model should yield a $\chi^2 \approx \nu$ \citep{2001MNRAS-Schwarz}. Thus, a desirable $\chi^2_{\text{red}}$ value should be low, typically ranging from 1 to 2, indicating a satisfactory and well-fitted model. The values of the best-fitting model parameters along with their corresponding uncertainties are presented in Table~\ref{tab:T2_results}. These uncertainties were determined by individually varying each parameter while keeping others fixed at their best-fitting values until $\chi^2_{\text{red}}$ reaches 2, thereby establishing parameter uncertainties. This approach results in a free parameter uncertainty of about 3$\sigma$ \citep{2001MNRAS-Schwarz}.

In this study, we modeled a total of five spectra, which offer broader spectral coverage and feature a greater number of emission lines. These spectra span the initial month following the outburst, specifically at epochs 1, 2, 3, 4, and 5, corresponding to days +10.00, +13.23, +17.92, +22.92, and +27.88, respectively. When selecting the spectra for modelling, we intentionally excluded the initial days after the outburst to ensure that photoionization remains the primary ionization mechanism \citep{2021oodward, 2021ATel14718Albanese, 2021ATel14747Page}. Among these five epochs, the first three pertain to the nebular phase, while the last two belong to the coronal phase.

Initially, we ran several one-component models in an attempt to replicate the observed spectra. However, we encountered difficulties in reproducing certain prominent lines in the spectra, such as [\ion{N}{ii}] 5669, 5680 \AA, and \ion{O}{i} 6366 \AA, which are clearly evident in the observed spectrum. Additionally, many other lines, including [\ion{Ne}{iv}] 4714, 4724 \AA, \ion{Fe}{ii} 4491 \AA, and nearly all Balmer lines, were underestimated by the one-component model. Similar limitations of the one-component model have been noted in several previous studies, such as V1974 Cyg \citep{2005ApJVanlandingham}, V4160 Sgr \citep{2007ApJSchwarz}, RS Oph 2006 \citep{2018MNRASMondal}, and V1280 Sco \citep{2022ApJPandey}. The primary reason behind these limitations in the one-component model lies in the heterogeneous density distribution within the Nova ejecta shell. It consists of clumpy material with high-density embedded in a more diffuse gas \citep{1993AJshore, 1995A&AParesce, 2022MNRASpandey, 2018MNRASMondal}. To address this issue, we employed a two-component model, based on the assumption that the ejecta comprises two density components: clumps and diffuse regions. The clump region is responsible for generating lines of low ionization, which constitute the majority of observed lines. On the other hand, the diffuse region generates highly ionized emission lines without affecting the clump contribution to the spectrum. For fitting each epoch's spectrum, we executed two distinct models with the two densities, and then scaled each model by the respective covering factor, ranging from 0 to 1. The resulting models were combined to produce the final model, incorporating contributions from both density regions.

\begin{figure*}
	\centering
	\includegraphics[scale=0.75]{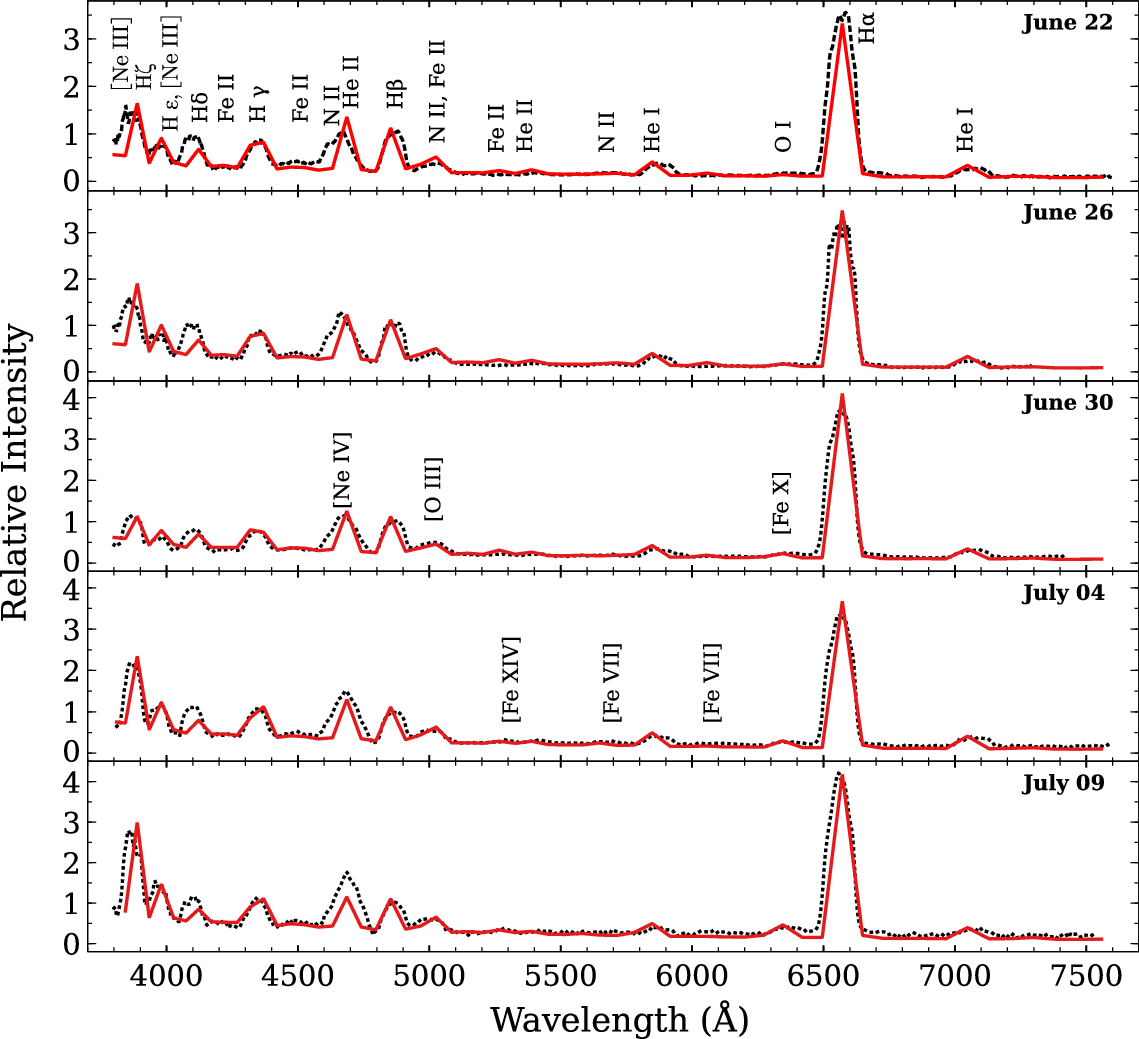}
	\caption[Best-fitting model plotted over the observed spectra.]{Best-fitting \textsc{cloudy} synthetic spectrum (red solid line) plotted over the observed spectrum (dashed black line) of V1674 Her obtained on 2021 June 23, June 27, June 30, July 4, and July 7, from top to bottom, respectively.
	}
	\label{fig:cloudy_best_fitting_model}
\end{figure*}

Fig.~\ref{fig:cloudy_best_fitting_model} illustrates the best-fitting \textsc{cloudy} model profiles (shown in red) superimposed on the observed optical spectra (shown in black) for five distinct epochs. Table \ref{tab:T3_chi_square} presents a comparison of relative fluxes between the best-fitting model-predicted lines and the observed lines during the early phase, along with corresponding $\chi^2$ values. For computing $\chi^2$ values, emission lines appearing in both observational and modelled spectra were selected. The observed and modelled line fluxes were determined using \textsc{IRAF}, and profiles with multiple components were decomposed using multiple Gaussian functions. To mitigate inaccuracies related to flux calibration across different epochs, flux ratios of observed and modelled emission lines relative to \ion{H}{$\beta$} were calculated.
\begin{table*}
	\centering
	\caption{Best fit \textsc{cloudy} model parameters during outburst phase of Nova Her 2021. }
	\label{tab:T2_results}
	\begin{tabular}{lllllr}
		\hline
		\multirow{2}{*}{Parameters}  &  \multicolumn{5}{c}{Values}\\
		\cline{2-6}
		&Epoch 1&Epoch 2&Epoch 3&Epoch 4&Epoch5\\
		\midrule
		Black Body Temperature ($\times 10^5$K)&1.99 $\pm$ 0.05& 2.04  $\pm $ 0.10&2.19 $\pm$ 0.05 &2.24 $\pm $ 0.10&2.34 $ \pm $ 0.10\\
		Luminosity ($\times 10^{38}$ \ergs)&1.26 $\pm$ 0.33 &1.99$_{\textrm{-0.41}}^{\textrm{+0.24}}$&2.00$_{\textrm{-0.41}}^{\textrm{+0.52}}$&3.16$_{\textrm{-0.34}}^{\textrm{+0.39}}$&3.16$_{\textrm{-0.28}}^{\textrm{+0.31}}$ \\
		Clump Hydrogen density ($\times 10^{8}\text{cm}^{-3}$)&10.0$_{\textrm{-1.19}}^{\textrm{+1.22}}$ &8.13$_{\textrm{-0.09}}^{\textrm{+0.19}}$&7.94$_{\textrm{-0.18}}^{\textrm{+0.19}}$&6.32$_{\textrm{-0.14}}^{\textrm{+0.15}}$& 5.62$_{\textrm{-0.38}}^{\textrm{+0.69}}$\\
		Diffuse Hydrogen density ($\times 10^{7}\text{cm}^{-3}$)&6.31$_{\textrm{-1.29}}^{\textrm{+1.63}}$ &3.98$_{\textrm{-1.03}}^{\textrm{+0.43}}$&3.16$_{\textrm{-0.34}}^{\textrm{+0.39}}$& 1.99$_{\textrm{-0.21}}^{\textrm{+0.24}}$ &1.00$_{\textrm{-0.05}}^{\textrm{+0.05}}$\\
		$\alpha$	&-3.00 &-3.00& -3.00 &-3.00&-3.00\\
		Inner radius ($\times 10^{14}\text{cm}$)$^a$	& 2.51& 3.98&4.68&5.75&7.08 \\
		Outer radius ($\times 10^{14}\text{cm}$)$^a$	&5.62 &7.94& 8.51&10.00& 15.14\\
		Filling Factor$^a$	& 0.10&0.10&0.10 & 0.10& 0.10\\
		$\beta$$^a$	& 0.00&0.00&0.00&0.00&0.00 \\
		Clump to diffuse Covering factor&62/38 &55/45 &53/47&53/47& 55/45 \\
		\ion{He}{}/\ion{He}{}$_{\sun}$&2.80$_{\textrm{-0.30}}^{\textrm{+0.70}}$ &2.50$_{\textrm{-0.20}}^{\textrm{+0.30}}$  &2.90$_{\textrm{-0.30}}^{\textrm{+0.50}}$ &2.90$_{\textrm{-0.50}}^{\textrm{+0.30}}$&$2.90^{+0.1}_{-0.2}$\\
		\ion{O}{}/\ion{O}{}$_{\sun}$&5.00$_{\textrm{-0.50}}^{\textrm{+0.50}}$ &5.50$_{\textrm{-1.00}}^{\textrm{+2.00}}$& 10.0 $\pm$ 2.5&6.00$_{\textrm{-2.00}}^{\textrm{+1.50}}$&8.00$_{\textrm{-0.20}}^{\textrm{+0.10}}$\\
		\ion{N}{}/\ion{N}{}$_{\sun}$&50.0$_{\textrm{-20.0}}^{\textrm{+30.0}}$ &50.0$_{\textrm{--5.00}}^{\textrm{+25.0}}$  &50.0$_{\textrm{-15.0}}^{\textrm{+20.0}}$ &30.0 $\pm$ 10&40.0$_{\textrm{-10.0}}^{\textrm{+3.00}}$\\
		\ion{Ne}{}/\ion{Ne}{}$_{\sun}$	&10.5$_{\textrm{-1.50}}^{\textrm{+2.00}}$ &15.5$_{\textrm{-1.50}}^{\textrm{+1.00}}$ & 12.5$_{\textrm{-2.50}}^{\textrm{+3.50}}$ &15.5$_{\textrm{-2.00}}^{\textrm{+4.00}}$&30.5$_{\textrm{-4.00}}^{\textrm{+5.00}}$ \\		
		Ejected matter mass ($\times 10^{-5}~M_{\sun}$) &0.867 &2.138 &2.918 &3.825&8.93\\
		Number of lines &16.0 &17.0 &15.0&16.0&14.0 \\
		Number of free parameters & 10.0& 10.0& 10.0&10.0&10.0\\
		Degrees of freedom &6.00 &7.00 &5.00 &6.00&4.00\\
		$\chi_{\text{tot}}^2$ & 10.91& 7.53&5.24&6.13&6.86\\
		$\chi_{\text{red}}^2$ &1.82&1.08&1.05 &1.02&1.71\\
		\hline
	\end{tabular}\\
	{\raggedright  Note:  $^a$ stands for the quantity is not considered as a free parameter.       \par}
\end{table*}

\subsubsection{Temperature and Luminosity}
\begin{figure}
	\begin{center}
		\begin{tabular}{c c}
		\includegraphics[scale=0.5]{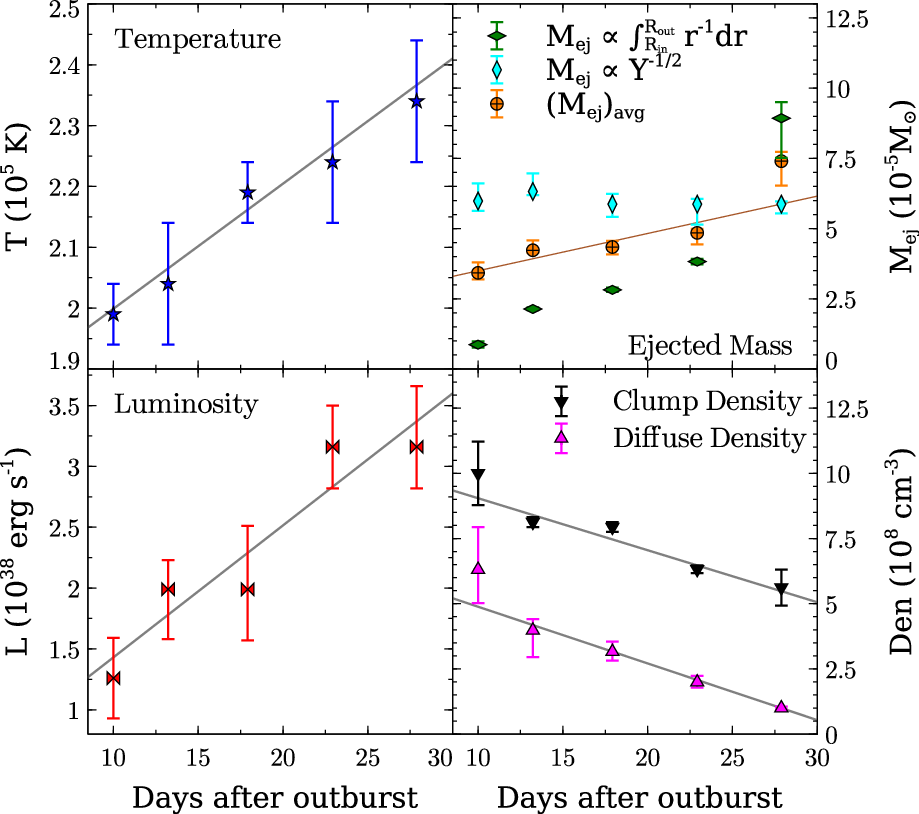}
		\end{tabular}
	\end{center}
	\caption{ Variations in the cloudy best-fitting parameters: temperature, luminosity, ejected mass, and number density. The panel for ejected mass comprises ejected masses calculated by two methods and their averages (for more information see \ref{em}) .
	}\label{fig:Variation_of_parameters}
\end{figure}

The parameter values as deduced from the best-fitting \textsc{cloudy} models, are provided in Table \ref{tab:T2_results}.  The optimal \textsc{cloudy} models indicate that the central ionizing source exhibited a temperature of $\sim$$1.99 \times 10^{5}$ K during epoch 1, which subsequently increased to around $2.34 \times 10^{5}$ K by epoch 5. Furthermore, the luminosity of the central ionizing source is found to be $1.26 \times 10^{38}$ \ergs~during epoch 1 and then raised to $3.16 \times 10^{38}$ \ergs~by epoch 5. These derived temperature and luminosity values from the best-fitting \textsc{cloudy} model reasonably align with the previously estimated values of T $\geq 4 \times 10^{5}$ K and L $\sim 1.0 \times 10^{38}$ \ergs) for day 16 \citep{2021RNAASRudy}. Both the effective temperature and luminosity of the central ionizing source exhibited lower values in the initial days and showed an increase in the subsequent epochs (refer to Fig.~\ref{fig:Variation_of_parameters}). This rise in the central ionizing source's temperature is due to the compression of leftover matter within the pseudo-photosphere as the ejecta expands during its collapse. This compression process results in elevated gas temperatures, consequently leading to higher rates of radiation emission \citep{1976MNRASBath}. The heightened radiation output subsequently triggers the expansion and cooling of the outer layers of the WD.

\subsubsection{Ejecta Density} \label{ed}
Throughout epochs 1 to 5, the diffuse hydrogen density varied from $6.31 \times 10^{7}$ to $1.00 \times 10^{7}$ $\text{cm}^{-3}$, while the clump hydrogen density ranged from $1.00 \times 10^{9}$ to $5.60 \times 10^{9}$ $\text{cm}^{-3}$. During the expansion phases of a nova outburst, the density of the ejected material undergoes a substantial decrease over time. This occurs as the ejecta expands at high velocities after an outburst, with some material being absorbed by interstellar clouds or the surrounding gas from the companion star. This absorption releases extra energy from the nova mechanism, causing a gradual shift towards an optically thin atmosphere and a subsequent reduction in density. The diminishing density exposes the ionizing radiation emitted by the central WD to the expanding ejecta, resulting in an increased ionization potential. This, in turn, facilitates the emergence of emission lines with higher ionization potentials. The density evolution observed in novae, along with the subsequent ionization processes, can generally be attributed to the interplay between expanding ejecta, matter absorption, and ionizing radiation from the central WD.

The ratio of clump-to-diffuse covering factors from the best-fitting model was calculated to be $62/38$, $55/45$, $53/47$, $53/47$, and $55/45$ for epochs 1, 2, 3, 4, and 5, respectively. We observed that the majority of emission lines originated within regions of higher ejecta density, as earlier predicted in Section \ref{lp}. The best-fitting \textsc{cloudy} model revealed that the \ion{He}{i} 5878 and 7065 \AA~ lines exclusively originated from the clump component of the ejecta. Conversely, the [\ion{N}{ii}] 5669, 5680 \AA, and \ion{O}{i} 6366 \AA~ lines were observed to be produced solely from the diffuse component of the ejecta. Thus, the modelling clearly illustrates that the observed spectra cannot be adequately replicated by a one-component model, and it is not possible to deduce the physical and chemical attributes of the source and ejecta solely from such spectra.

\subsubsection{Elemental Composition} \label{ec} 
During epochs 1, 2, and 3 (nebular phases), oxygen, nitrogen, iron, helium, and neon were found to be overabundant relative to the solar value. With the exception of iron, whose abundance decreased to a solar value during epochs 4 and 5 (coronal phases), these elements continued to be overabundant. Furthermore, the model suggested a twofold increase in neon abundance within the initial days of the coronal phase. The notable increase in \ion{Ne}{} abundance and decrease in \ion{Fe}{} abundance could be attributed to Nova V1674 Her's shift from a \ion{Fe}{ii} type to a Neon-type nova, as suggested earlier by \citet{2021ATel14746, 2022RNAASWoodward}. Moreover, the neon overabundance in the ejecta indicates that the WD was an ONe type as predicted in Section \ref{wdm} \citep{2012BASIEvans}.

Helium abundance across all epochs was determined by fitting the prominent \ion{He}{i} (5876, 7065 \AA) and \ion{He}{ii} (4686, 5412 \AA) lines. During modelling, we observed that the \ion{He}{i} lines originated from high-density regions, while \ion{He}{ii} lines arose from both higher and lower densities. Oxygen abundance in epoch 1  was determined using \ion{O}{i} 6366 \AA~ alone, while in subsequent epochs, \ion{O}{iii} 5007 \AA~ was also considered. \ion{O}{i} originated from a lower density zone, whereas [\ion{O}{iii}] emerged from higher density areas. Nitrogen abundance for all three epochs was estimated through fitting \ion{N}{ii} 5680 \AA, which originated from the lower density region. Neon abundance for epochs 1 and 2 was derived from [\ion{Ne}{iii}] (3869, 3968 \AA), and for epochs 3, 4, and 5, the [\ion{Ne}{iv}] 4714 \AA~ line was included. Both [\ion{Ne}{iii}] and [\ion{Ne}{iv}] primarily originated from the clump region. Iron abundance in epochs 1, 2, and 3 was determined by fitting \ion{Fe}{ii} (4200, 4487, 5018, 5168, 6247 \AA), with epoch 3 contributing from the coronal line [\ion{Fe}{x}] 6375 \AA. These lines were mainly generated from higher density regions, although the lower density area also contributed. Enhanced iron abundance over the solar value during the first three epochs implied helium enrichment in the nova ejecta, potentially indicating that the companion star had evolved beyond the main-sequence level \citep{2008Bodebook}. For epochs 4 and 5, iron abundance was measured by fitting [\ion{Fe}{x}] 6375 \AA, [\ion{Fe}{vii}] 6087 \AA, [\ion{Fe}{vi}] 5677 \AA, [\ion{Fe}{xiv}] 5303 \AA, \ion{Fe}{ii} 5167 \AA, and \ion{Fe}{ii} 4487 \AA.

 A notable increase in intensity was observed in emission features such as \ion{H}{$\eta$} 3835.4 \AA, \ion{H}{$\zeta$} 3889 \AA, and \ion{H}{$\epsilon$} 3970 \AA~ starting on day 10 (2021 June 22.99). Similarly, the \ion{Fe}{ii} 5018 \AA~ emission feature and an emission line identified as a blended feature of \ion{N}{ii} 4638 \AA~ and \ion{He}{ii} 4686 \AA~ exhibited similar behaviour on day 17.9 (2021 June 30). By varying the hydrogen density, effective temperature, and luminosity of the source, we could not generate these features in the model spectra, to match the observed one. These characteristics were exclusively generated by enhancing the abundances of neon and oxygen beyond their solar values \citep{2010ApGrevesse}. Consequently, the emission features of [\ion{Ne}{iii}] 3869, 3968 \AA, [\ion{Ne}{iv}] 4714, 4724 \AA, and [\ion{O}{iii}] 5007 \AA~ became stronger, and the intensity of the blended  features at $\lambda = 3889, 4686,~\text{and}~ 5018$ \AA~ increased. This result strongly supports the claim we made in Section \ref{nebu} that the nebular phase of Nova V1674 Her began on day 10, marking one of the quickest transitions in history.

\subsubsection{Ejected mass Calculation} \label{em}
The ejected mass within the model shell could be calculated using the following equation, \citep{2001MNRAS-Schwarz}:
\begin{equation}
	M_{\text{ej}} = n(r_0) f(r_0) \int_{R_{\text{in}}}^{R_{\text{out}}} \left(\frac{r}{r_0}\right)^{\alpha+\beta} 4\upi r^2 dr,
\end{equation}
where $n(r_0)$  represents the hydrogen density ($\text{cm}^{-3}$) and $f(r_0)$ stands for the filling factor at the inner radius of the shell ($r_0$). The exponents $\alpha$ and $\beta$ correspond to the power laws. 
The values for density, filling factor, $\alpha$, and $\beta$ are directly adopted from the best-fitting \textsc{cloudy} model parameters (refer to Table \ref{tab:T2_results}).  The estimation of the total ejected shell mass involved multiplying the mass in both density components (clump and diffuse) with their corresponding covering factors and subsequently adding them together. Consequently, the ejected hydrogen shell masses for epochs 1, 2, 3, 4, and 5 are estimated to be: $0.87\times 10^{-5} M_{\odot}$, $2.14 \times 10^{-5} M_{\odot}$, $2.92 \times 10^{-5} M_{\odot}$, $3.83 \times 10^{-5} M_{\odot}$, and $8.93 \times 10^{-5} M_{\odot}$, respectively. The ejected mass could also be estimated from the helium abundance factor $\text{Y}$ \citep{1993AJshore}, 
\begin{equation}\label{shore}
	M_{\text{ej}} = \text{Y}^{-1/2}~10^{-4}~M_{\sun},
\end{equation}
where $\text{Y}$ represents the helium abundance enhancement factor, which was adopted from our \textsc{cloudy} best-fitting model parameters (Table \ref{tab:T3_chi_square}). 
This approach has been successfully applied to various novae, including neon novae like QU Vul \citet{1993AJshore} and V1974 Cyg \citep{2005ApJVanlandingham}. Based on this method, the estimated mass of the ejecta shell for epochs 1, 2, 3, 4, and 5 are found to be: $\sim$$5.98 \times 10^{-5}~M_{\sun}$, $6.33 \times 10^{-5}~M_{\sun}$, $5.87 \times 10^{-5}~M_{\sun}$, $5.87 \times 10^{-5}~M_{\sun}$, and $5.87 \times 10^{-5}~M_{\sun}$, respectively.

\begin{landscape}
	\begin{table}
		\centering
		\caption{Observed and best-fit \textsc{cloudy} model line flux values for nebular and coronal phase epochs.}
		\label{tab:T3_chi_square}
		\begin{tabular}{llccccccccccccccc}
			\hline
			\multicolumn{2}{c}{}&\multicolumn{3}{c}{Epoch 1 (June 22)}&\multicolumn{3}{c}{Epoch 2 (June 26)}&\multicolumn{3}{c}{Epoch 3 (June 30)}&\multicolumn{3}{c}{Epoch 4 (July 04)}&\multicolumn{3}{c}{Epoch 5 (July 09)}\\
			\cline{3-17}
			Line ID & $\lambda$ ({\AA}) &modelled & Observed & $\chi^2$ & modelled& Observed & $\chi^2$ & modelled& Observed & $\chi^2$ &modelled & Observed & $\chi^2$ &modelled  & Observed& $\chi^2$ \\
			\hline
			\ion{H}{$\zeta$},{[\ion{Ne}{iii}]} & 3889 & 1.519 & 1.748 &  2.339 &1.508 & 1.567 & 0.152 & 1.195  & 1.114 &  0.292 & 1.909  & 1.90&  0.002 & 2.138  & 2.485 &  5.352  \\
			\ion{H}{$\epsilon$},{[\ion{Ne}{iii}]} & 3970 & 1.080 & 0.964 &  0.600&1.106  & 1.013 &0.381 & 0.852  & 0.747 &   0.491 & 1.288  & 1.089 &  1.754 & 1.438  & 1.525 &  0.342  \\
			\ion{H}{$\delta$} & 4101 & 0.785 & 0.921 &  0.814&0.867  & 0.879 &  0.007 & 0.805  & 0.770 &  0.053 & 1.124  & 1.077 &  0.095 & 1.059  & 1.143 &  0.31892  \\
			\ion{Fe}{ii},[\ion{Fe}{ii}] & 4180 & 0.653 & 0.594 &  0.153&0.800  & 0.700 &  0.444 & - & - &  - & - & - &  - & - & - &  -  \\
			\ion{H}{$\gamma$} & 4340 & 1.266 & 1.047 & 2.145& 1.204 &1.033  & 1.288 & 1.239  & 1.015 &2.223 & 1.469  & 1.358 &   0.546 &1.537   &1.415   & 0.669   \\
			\ion{Fe}{ii} & 4415 & 0.806 & 0.745 & 0.166&0.860   &0.821  & 0.067 & 0.715 & 0.669  & 0.097 &1.049    &0.937  &  0.564  &1.292  & 1.291  &  0.2E-4 \\
			\ion{N}{iii}&4638  & 0.512 &0.804  &3.793&0.694   &0.654  &0.069 &-  & -  &- & -  & - & - &-   &-  &-  \\
			\ion{He}{ii}&4686  & 0.745 &0.729  &0.010&0.659   &0.984  &4.712 &0.919   &1.097  &1.397 & 0.642   & 0.878 & 2.481 &1.009    &1.014  &0.0013 \\
			{[\ion{Ne}{iv}]}&4724  & - &-  &-&-  &-  &- &0.192   & 0.117  &0.249 & 0.383   & 0.445 & 0.172 &-   &-  &-  \\
			\ion{H}{$\beta$} &4861  & 1.000 & 1.000 & 0.000& 1.000 & 1.000 & 0.000 & 1.000 & 1.000 & 0.000 & 1.000 & 1.000 & 0.000 & 1.000 & 1.000 & 0.000\\
			{[\ion{O}{iii}]} & 5007 & - & - &-& - & - & - &0.768   &0.699  &0.210  &0.896  &0.807   & 0.351 &0.897  &0.935 &  0.063  \\
			\ion{Fe}{ii},\ion{He}{i} & 5016 & 0.839 & 0.716 &0.671& 0.671  & 0.644& 0.030 & -  &-  &- &-   &-  & - &-  &-  & -  \\
			\ion{Fe}{ii}  &5276  &0.274 & 0.246 &0.035& 0.304  & 0.224 & 0.278 & 0.335   &0.280  & 0.135 & -  & - & - &-   &-  & -  \\
			{[\ion{Fe}{xiv}]}  &5303  &- & - &-& - & - & - & -  &-  & -& 0.427  & 0.407 & 0.016  &0.396  &0.418  &  0.023  \\
			\ion{He}{ii}  &5412  &0.264   & 0.242 &0.021& 0.282  & 0.279 & 0.0004 & 0.267   &0.240  & 0.031 & 0.326   & 0.304 & 0.021 &0.351   &0.374  & 0.024  \\
			{[\ion{Fe}{vi}]}  &5677  &- & - &-& - & - & - & -  &-  & - & 0.243   & 0.217 & 0.031 &0.286    &0.281  & 0.0011  \\
			\ion{N}{ii}  &5679  &0.339   & 0.325 &0.009& 0.322  & 0.298 &  0.027 & -  &-  & - & -  & - & - &-   &-  & -  \\
			\ion{He}{i} &5875  & 0.423 &0.384 &0.066 & 0.367  & 0.333 & 0.051  &0.359    &0.331  & 0.034& 0.356   & 0.324 &0.046 &0.454   &0.422   &   0.046 \\
			\ion{Fe}{ii} & 6248 & - & -& - & 0.164  & 0.180 &0.011  &-    &-  & - & -  & - &- &-  & -  &   - \\
			\ion{O}{I} & 6366 &0.223  &0.233 & 0.005 & 0.164  & 0.164 &0.4E-4  &-   &-  & - & -  & - &- &-  & -  &   - \\
			{[\ion{Fe}{vii}]} &6087  &- &- & - & - & - &-  &0.084   &0.069  & 0.008 & 0.121   & 0.123 &1.7E-4 &-  & -  &   - \\
			{[\ion{Fe}{x}]} &6375  &- &- & - & - & - &-  &0.245    &0.232  & 0.008 & 0.180  & 0.155 &0.029 &0.371   & 0.370  &   0.3E-4 \\
			\ion{H}{$\alpha$}  &6562  & 3.453 &4.429  &10.906$^*$ & 4.983 &4.248  &13.494$^*$  &3.179   &2.706  &5.605$^*$  & 2.806  & 2.048 &14.351$^*$ &2.936   &2.546 & 3.812$^*$ \\
			\ion{He}{i}&7002  &0.369  &0.326  & 0.082& 0.282  &0.296  & 0.008 &0.299   & 0.283 & 0.0114 &0.321    & 0.298 &  0.024& 0.251  & 0.232 &0.016 \\
			\hline	
		\end{tabular}
		{\raggedright  Note:  The symbol  $^*$ denotes that the value is not taken into account while calculating the total $\chi^2$ for each epoch.       \par}
	\end{table}
\end{landscape}

We chose to use the average values of the ejected masses obtained from the two aforementioned approaches as the most plausible estimates for the expelled masses during epochs 1, 2, 3, 4, and 5, which are approximately $3.42 \times 10^{-5}~M_{\sun}$, $4.23 \times 10^{-5}~M_{\sun}$, $4.35 \times 10^{-5}~M_{\sun}$, $4.85 \times 10^{-5}~M_{\sun}$, and $7.40 \times 10^{-5}~M_{\sun}$, respectively.
Given the considerable fluctuations in ejected mass observed across the five distinct epochs, particularly in the first and last epochs, adopting the average is considered a more reliable estimation. These estimates of ejected mass across the five distinct epochs are consistent with the typical expectations for a neon nova, which ejects a smaller mass ($\sim 10^{-6} - 10^{-5}~M_{\sun}$) \citep{gehrz_2008, 1998PASPGehrz, duerbeck_2008}. However, it is important to note that the obtained values of ejected mass could be considered as their minimum value because a large amount of neutral or much hotter gas might remain undetected or require a more intricate approach to be detected \citep{1998ASPCFerland, 2008Bodebook}.

TNRs on massive WDs, like V1674 Her, lose their envelopes and fade more swiftly than those on small WDs (CO type) \citep{2018ApJShara}. Those novae with higher WD mass and faster ejection velocities release less mass ($\sim 10^{-6} - 10^{-5}~M_{\sun}$) than lower WD mass and slower speed novae \citep{gehrz_2008, 1998PASPGehrz, duerbeck_2008}. As a result, novae of ONe type eject smaller amounts of material than CO novae. A massive WD requires comparatively less accreted matter to trigger a TNR \citep{1998PASPGehrz, 1994AJShara}, a concept reflected in the mass-radius relationship ($M_{\text{WD}}^{1/3}R_{\text{WD}} = \text{constant}$) and the binding energy formula $(GM_{\text{WD}}~M_{\text{ej}})/R_{\text{WD}}\propto M_{\text{WD}}^{4/3}~M_{\text{ej}}$. \citet{2021oodward} suggested a maximum possible ejected mass of $\sim$$\sim 1.4^{+0.8}_{-1.2} \times 10^{-3} M{\sun}$ for Nova V1674 Her. Conversely, \citet{2021ApJDrake} predicted an ejection mass range of $2 \times 10^{-5}~M_{\sun}$ to $2 \times 10^{-4}~M_{\sun}$ for such a rapid nova. This study estimates ejected masses at five epochs between $3.42$ and $7.40 \times 10^{-5}~M_{\sun}$, reasonably consistent with prior research. Additionally, the central WD mass is estimated at $\sim~1.36~M_{\sun}$, also in line with previous studies.

\subsection{Morpho-kinematic modelling}\label{mkm}
We performed a morpho-kinematical analysis of the outburst of Nova V1674 Her in 2021, using the latest version of {\footnotesize \textsc{SHAPE}}\footnote{\url{https://wsteffen75.wixsite.com/website/}} (i.e., \textsc{shapex}; \citet{2017AC_Steffen, 2012Steffen}). Our aim is to generate the 3-D structure of the ejecta by modelling prominent emission lines in the observed spectra and studying how the structure evolves with time. Through morpho-kinematic analysis, one can determine the inclination angle, position angle, geometrical size, and structure of the ejecta shells. Similar studies are conducted for a few other novae, \citep[for example,][etc]{2011MNRAS410Munari,2020MNRASHarvey,Linford2015, 2020MNRASPavana}.

\subsubsection{Gaussian decomposition of \ion{H}{$\alpha$} line}
\begin{figure*}
\centering
\includegraphics[scale=0.6]{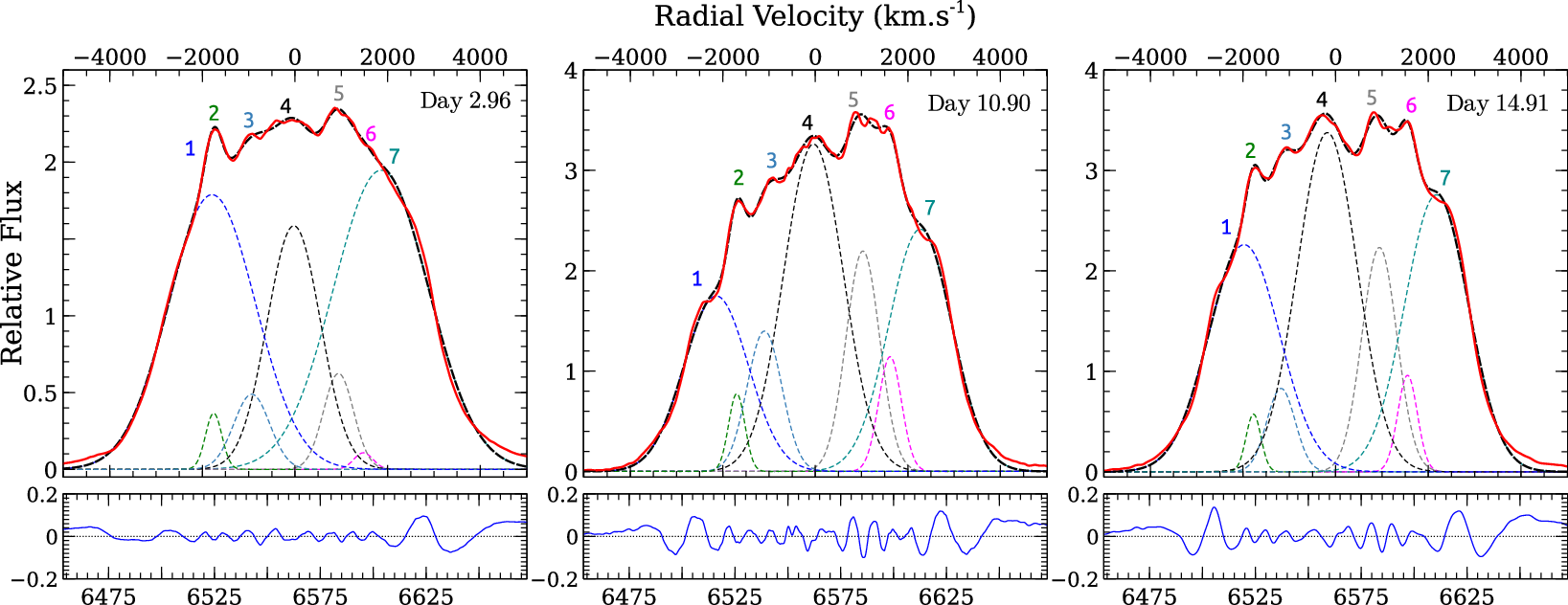}\\
\vspace{0.4cm}
\includegraphics[scale=0.6]{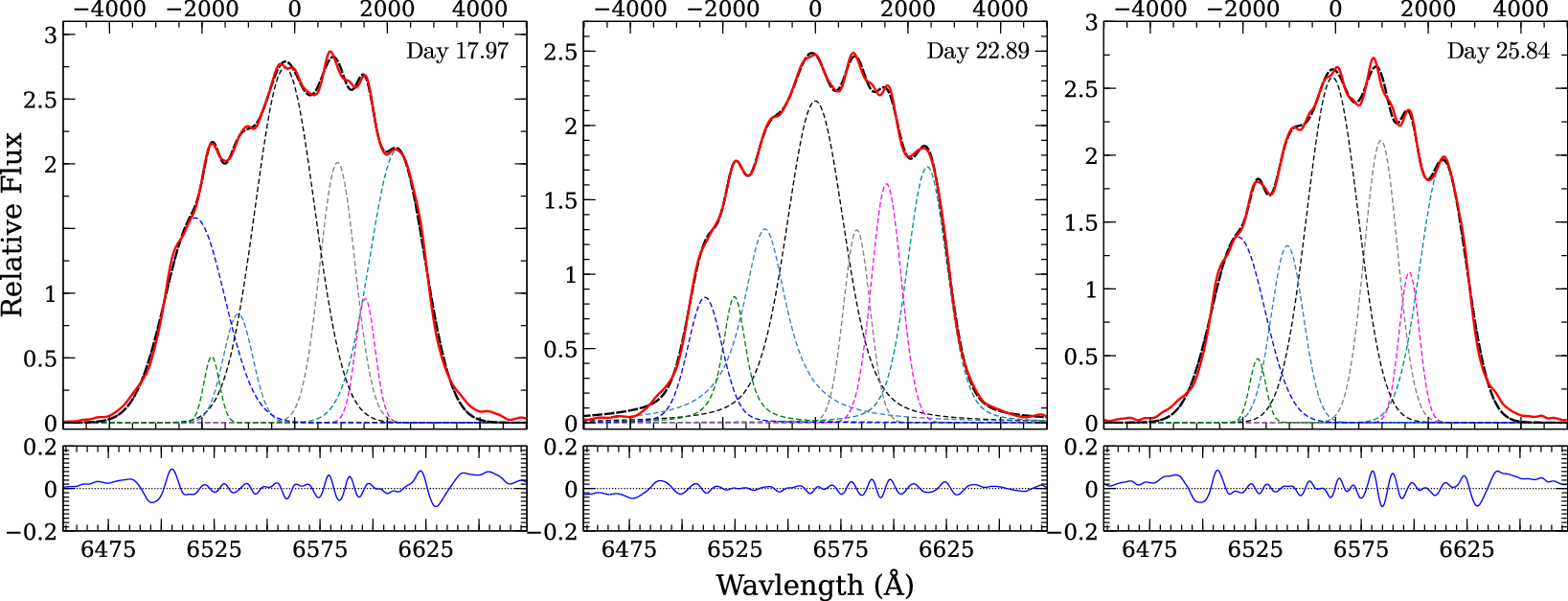} 
\caption{The best fit for the Gaussian components of the \ion{H}{$\alpha$} line profile of V1674 Her on days 2.96, 10.897, 14.91, 17.972, 22.886, and 25.838 following the outburst. The combined Gaussian components are depicted with black broken lines, while the observed spectrum are represented by red solid lines. Each individual Gaussian component is distinguished by a unique colour scheme shown in each panel. Component numbers 1 through 7 correspond to peaks listed in increasing wavelength order. Additionally, the residuals from the overall fit are displayed at the bottom of each panel.	
}
\label{fig:Ha_gaussian_fitting}
\end{figure*}

Due to the remarkably high ejection velocity of the nova, the majority of line profiles exhibit broad characteristics, accompanied by a series of ripples on top of each profile (see Fig.~\ref{fig:Ha_gaussian_fitting}). To identify components with different velocities, we performed Gaussian fitting on \ion{H}{$\alpha$} emission lines using the {\textsc{fityk}}\footnote{\url{https://fityk.nieto.pl/}} software \citep{WojdyrFityk}. It is a customizable peak-fitting application with a user-friendly graphical interface, designed to manage a wide variety of non-linear functions, and enables users to perform continuum subtraction and Levenberg-Marquardt least squares fitting.

We fitted the emission line profiles of \ion{H}{$\alpha$} observed on 2021 June 15.945 (2.96 d), June 23.856 (10.897 d), June 27.867 (14.91 d), June 30.931 (17.972 d), July 04.914 (22.886 d), and July 07.866 (25.838 d) UT, spanning over the first month. \ion{H}{$\alpha$} was chosen to be modelled because it stands out as the broadest and most distinct line across all time frames. To fit the observed line profiles, we required at least seven Gaussian components for each epoch (see Fig.~\ref{fig:Ha_gaussian_fitting}). This enabled us to obtain the radial velocities and the \textsc{FWHM} values of the individual components. Reasonable velocity symmetry appears to exist between the fast-moving components, 1 and 7, as well as the slower-moving components, 2 and 6, and 3 and 5. Table~\ref{tab:Radial_velocity} and \ref{tab:FWHM_velocity} provides the radial and \textsc{FWHM} velocities of each ripples. All profiles exhibit good correspondence in radial velocity for the same ripple, indicating that they are kinematically identical. These kinds of ripples occur due to various circumstances: (1) when novae eject large, distinct blobs of matter at various angles to the line of sight \citep{2006A&Amunari}, (2) when the density distribution appears in-homogeneous and speed variation results from it \citep{2007ApJSchwarz}, and (3) when brightness increases as a result of the polar and projection effects of equatorial rings \citep[][and references therein]{2006A&Amunari}.

\begin{table*}
\centering
\caption{Calculated radial velocity of the ripples numbered on the \ion{H}{$\alpha$}  emission line profile in V1674 Her.}
\label{tab:Radial_velocity}
\begin{tabular}{llcccccc}
\hline
\multirow{2}{*}{}  &  \multicolumn{7}{c}{Component Radial velocity (\kms )}\\
\cline{2-8}
Day& 1&2&3&4&5&6&7\\
\hline
2.96  &-1791.20&-1755.56&-933.19&-24.12&940.36&1477.00&1858.72\\
10.90&-2141.98&-1698.25&-1098.42&-39.36&1028.17&1616.80&2270.14\\

14.91 &-1946.50&-1774.17&-1180.35&-180.07&943.69&1550.36&2176.99\\

17.97&-2151.20&-1793.28&-1222.92&-211.49&923.76&1522.17&2213.57\\

22.89&-2385.42&-1751.07&1091.41&-0.74&893&1535.41&2416.47\\

25.84 &-2099.2&-1681.79&-1042.08&67.60&978.70&1590.10&2323.21\\
\hline
\end{tabular}
\end{table*}

\begin{table*}
\centering
\caption{ Calculated \textsc{FWHM} velocity  of the ripples numbered on the \ion{H}{$\alpha$}  emission line profiles in V1674 Her.}
\label{tab:FWHM_velocity}
\begin{tabular}{llcccccc}
\hline
\multirow{2}{*}{}  &  \multicolumn{7}{c}{Component \textsc{FWHM} velocity (\kms )}\\
\cline{2-8}
Day& 1&2&3&4&5&6&7\\
\hline
2.96 &2205.21&418.14&914.73&1355.92&767.73&486.29&2412.79\\
10.90&1632.34&415.02&884.39&1518.62&859.59&584.94&1605.69\\

14.91 &1804.5&378.35&723.22&1518.31&893.25&499.43&1617.06\\

17.97&1525.48&393.76&746.32&1511.60&895.77&498.56&1391.97\\

22.89&729.04&164.66&1144.26&1475.42&672.94&751.66&1031.64\\

25.84 &1351.17&402.03&819.33&1302.66&852&511.85&1170.44\\
\hline
\end{tabular}
\end{table*}

\subsubsection{SHAPE modelling}
We estimated the 3D morphological structure and density distribution in the ejecta of Nova V1674 Her by modelling the prominent emission line profiles. Line profiles from the expanding ejecta provide insights into how morphology and kinematic properties correlate along the line of sight. We can observe this correlation in either 1D or 2D spectral profiles, enabling us to attain the position-velocity (PV) profiles. These profiles usually encompass both the position and velocity of a specific point across the shell ejecta.

Within \textsc{shape}, we possessed the capability to construct diverse 3D multi-component structures, including polar blobs, prolate structures with tropical rings, prolate structures exhibiting equatorial rings, and bipolar structures. These structures were constructed using a variety of elementary geometries, such as spheres, cylinders, cones, or toroids, either individually or in combination. Moreover, the fundamental geometry can be tailored by employing various structural modifiers like squeeze, spiral, twist, and others. Each component was assigned a density profile and velocity field using corresponding modifiers. The \textsc{shape} code conducts all the radiation transfer calculations through the ejecta and generated the 3D model, along with the PV diagram.

In this study, we reconstructed the 3D structure of the nova ejecta using the \ion{H}{$\alpha$}-PV spectral profile of the ejecta. The line profiles of \ion{H}{$\alpha$} generally provided a well-resolved view of the position-velocity correlation characteristics towards the line of sight. Our main objective is to develop a 3D model characterized by a designated density distribution and expansion velocity field, ensuring that the PV profile generated by the model corresponded with the observed position-velocity profile. To achieve this, we utilised fundamental modifiers related to velocity, density, squeeze, and translation, which offered additional characterization for each constituent. Through the utilization of these modifiers, we adjusted all unconstrained variables to generate synthetic line profiles that closely resembled the observed ones. One important modifier in the \textsc{shape} software package is the 'squeeze' modifier, as defined by \citep{2013ApJRibeiro}:
\begin{equation}
    \text{squeeze}=1-\frac{a}{b},
\end{equation}
where $a$ and $b$ represent the semi-minor and semi-major axes of the ejecta shell, respectively. This equation was utilised to transform a spherical shell configuration into a bipolar structure. A higher value of the squeeze indicates a more pronounced deformation at the ejecta's waist, while a squeeze value of zero implies a spherical shape. In this study, we adjusted the values of the squeeze modifier within the range of 0.0 to 1.0. Translation modifiers were employed to reposition the polar caps along with their corresponding coordinate systems to their respective poles. The study encompasses the modification of both the parameters of the modifiers and the inclination and position angles of the binary system. These angles ranges from 0 to 90$^{\circ}$ and 0 to 180$^{\circ}$, respectively. To determine the optimal model, various configurations were computed prior to discovery. The same radius-density power-law relation as in photoionization modelling, i.e., $\rho\propto r^{-3}$ (Section~\ref{pma}), was employed. The initial density guess was adopted from the best-fitting models produced by \textsc{cloudy} (see Table~\ref{tab:T2_results}). The determination of the radius for each component was based on the observed line profile's \textsc{FWHM} and the elapsed time since the outburst. To portray the radial velocity distribution in all directions, we employed the Hubble-flow type linear velocity field equation given by \citep{2009ApJRibeiro}:
\begin{equation}
\label{hub}
v(r)=\dfrac{3200}{\sin i}(\dfrac{r}{r_0}),
\end{equation}
where $i$, $r_0$, and $r$ represent the inclination angle, inner, and outer radii of the ejected shell. In all epochs, the shell's radius was calculated based on the expansion velocities. Previous studies \citep{2021ATel14723Woodward,2021ApJDrake} estimated that the maximum expansion velocity could reach up to 11,000 \kms. In this study, we explored the parameter space by varying the expansion velocity ($V_{exp}$) in the range of 1,000 to 11,000 \kms, with increments of 100 \kms.

Initially, a comparison between the observed and model images, as well as the PV diagrams, was conducted through visual estimation. Once a satisfactory visual alignment between the observed and modelled line profiles was achieved, the goodness of fit was assessed using two quantitative measures: the root mean square ($rms$) and the quality ($q$) factor. The $rms$ value was computed as a metric to quantify the overall differences between the synthetic and observed profiles, providing an estimation of the mean disparity between the two. On the other hand, the $q$-factor served as a measure of similarity based on the relative positions and strengths of the features present in the line profile. The $rms$ was calculated using the following equation:
\begin{equation}\label{rms}
rms=\sqrt{\frac{1}{n}\sum_{i}^{}\left[1-\frac{M_i}{O_i}\right]^2},
\end{equation}
where $M_i$ and $O_i$ represent the modelled and observed flux values of the $i^{th}$ observable, respectively. In this case, we focused on a single observable per model, so $i$ is equal to 1, resulting in a total number of observables ($n$) also being unity. A model is considered acceptable if the root mean square (rms) is less than one \citep{2020MNRASBandyopadhyay}. Additionally, to assess the quality of fitting, we computed the quality factor $q$, given by \citep{2009AAMorisset}.
\begin{equation}\label{eqqi}
q=\frac{\log(\frac{M}{O})}{\tau}, \qquad \text{where,} \quad \tau=\log(1+\delta),
\end{equation}
where $\delta$ represents the corresponding error between the modelled and observed line profiles. Similarly, the fitting process is deemed successful within the specified tolerance limit if the absolute value of $q$ is less than 1. Consequently, upon closer examination, we determined that a 10\% discrepancy could be attributed to inaccuracies in the model parameters.

We reconstructed the 3D ejecta geometry of V1674 Her for six epochs: days 2.96, 10.91, 14.91, 17.97, 22.89, and 25.84 following the outburst. Across all epochs, the modelled ejecta consistently exhibit a bipolar geometry characterised by a central equatorial ring surrounding the major axis. The bipolar structure consists of symmetrical bipolar morphology resulting from a squeezed oblate shape and asymmetrical polar caps. The presence of an equatorial ring in the \ion{H}{$\alpha$} morphology suggests potential interaction between the outflow and a secondary component. This phenomenon was observed in other previously analysed novae, such as V2672 Oph (Nova Oph 2009) \citep{2011MNRAS410Munari, 2011MNRASRibeiro}, and nova ASASSN-18fv \citep{2020MNRASPavana}.

\begin{figure}
  \centering
  \includegraphics[scale=0.46]{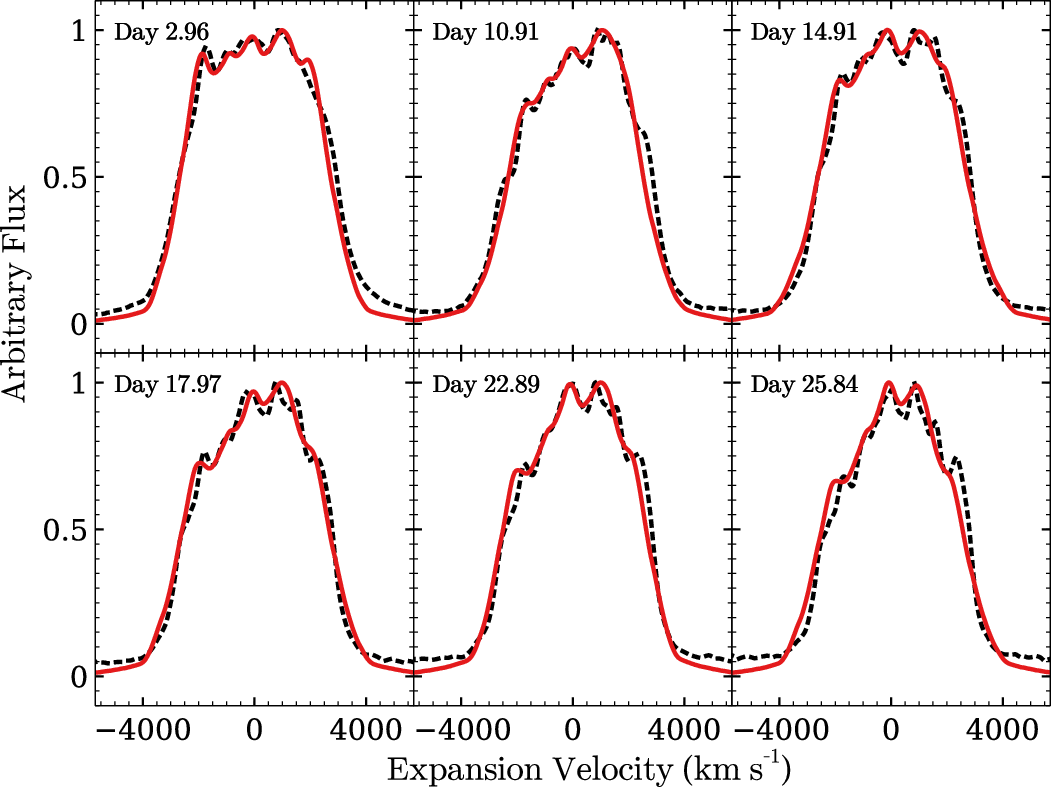}
  \caption{ Best-fit modelled velocity profile (the red solid line) overlaid on the observed \ion{H}{$\alpha$} profile (dashed black line) of V1674 Her on the following dates in 2021: June 15.95 (2.96 d), June 23.86 (10.91 d), June 27.87 (14.91 d), June 30.931 (17.97 d), July 04.91 (22.89 d), and July 07.87 (25.84 d).
  }
  \label{fig:havp}
  \end{figure}

Despite the reconstructed geometries exhibiting similarities across epochs, notable variations were identified in most of the free variables, including density, velocity, and radius. The inclination angle ($i$) and position angle ($PA$) experienced changes within a limited range. For all the modelled line profiles, we determine $i = 67 \pm 1.5^{\circ}$ as the optimal value for the ejecta morphology's inclination, while the position angle is $PA=35^{+20}_{-15}$. Fig. \ref{fig:havp} displays the best-fitting model profiles of \ion{H}{$\alpha$} overlaid on the observed line profiles. To assess the model's appropriateness, we calculated and provided the rms and $q$-factor values in Table \ref{tab:shape2} using equations (\ref{rms}) and (\ref{eqqi}).
\begin{table}
\centering
\caption{Comparison between observed and modelled line fluxes of V1674 Her.}
\label{tab:shape2}
\begin{tabular}{llcccc}
\hline
Day& O. Flux&M. Flux&Mod./Obs.&rms&$q$\\
\hline
2.96 &110.0&118.0&1.0727&0.0727&0.7366\\
10.90 &103.0&103.7&1.0068&0.0068&0.0711\\

14.91 &110.1&117.3&1.0065&0.0654&0.6646\\

17.97&102.4&108.8&1.0625&0.0625&0.6361\\

22.89&100.2&109.0&1.0878&0.0878&0.6758\\

25.84 &93.12&99.2&1.0653&0.0653&0.6636\\
\hline
\end{tabular}
\end{table}

  \begin{figure*}
  \centering
  \includegraphics[scale=0.68]{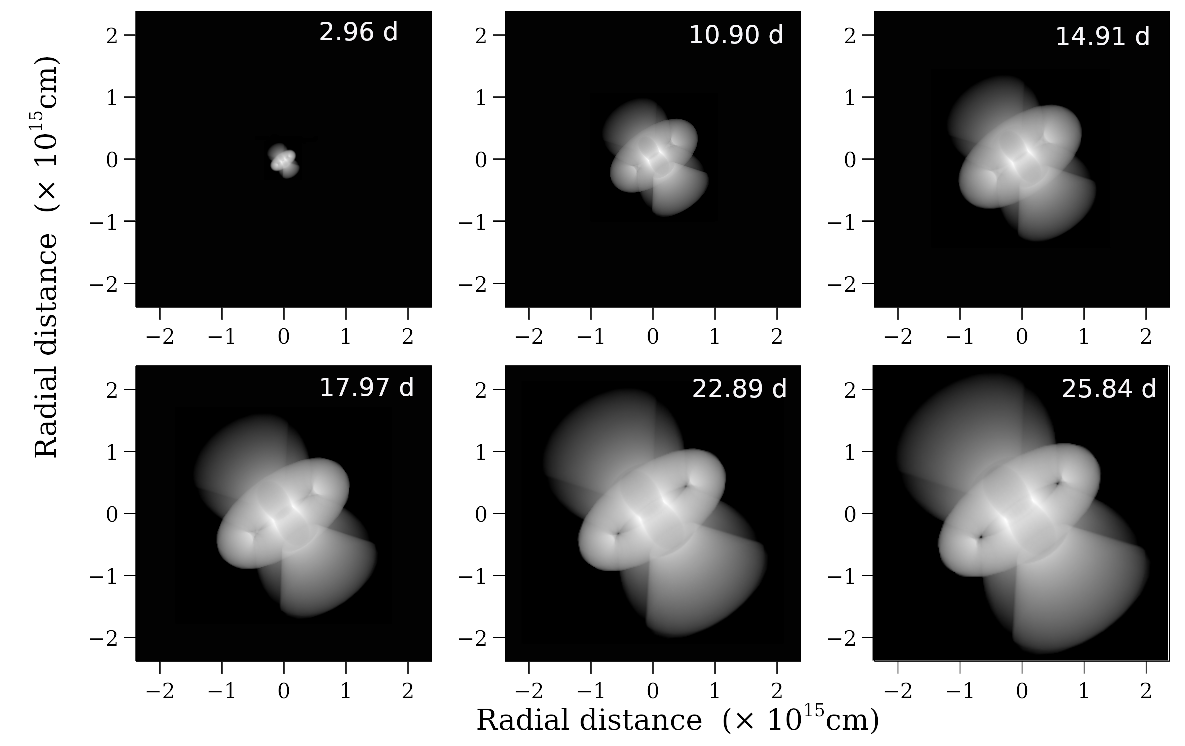}
  \caption{ 2D representation of the asymmetric 3D model result, obtained for the \ion{H}{$\alpha$} geometry of the ejecta of V1674 Her. The model images of the ejecta are derived from the best-fit to the \ion{H}{$\alpha$} line profile on the following dates in 2021: June 15.95 (2.96 d), June 23.86 (10.90 d), June 27.87 (14.91 d), June 30.931 (17.97 d), July 04.91 (22.89 d), and July 07.87 (25.84 d).
  }
  \label{fig:3d}
  \end{figure*}

The 3D configuration of \ion{H}{$\alpha$} within the V1674 Her ejecta is depicted in Fig.~\ref{fig:3d}. This figure illustrates the density variations within the ejected material shell along both the bipolar axis (Z-axis) and the equatorial ring axis (X-axis). According to the model, the density in the equatorial region of the ejecta shell is higher than that of the polar components on the specified days, which is evident in Fig.~\ref{fig:3d} where the denser areas are represented in white. The graphical representation also visualises the gradual expansion of the shell's geometrical dimensions during the intervals between the six epochs. The asymmetric features observed along the bipolar axis arise from the polar caps, and they are common among all model geometries within the context of the entire ejecta model. This asymmetry is attributed to unconstrained factors such as ejecta density, velocity, and angular radii within the polar cap regions. The red component of the polar caps exhibits a significantly higher magnitude compared to the blue component, with respective ratios of $\sim$3.11, 4.88, 3.88, 4.8, 4.49, and 3.13. This comparison is based not solely on geometric magnitude but also considers the density differences between the two poles. It indicates a larger amount of matter moving away from the line of sight (towards the red region) compared to that moving towards the line of sight (towards the blue region). With the exception of the polar cap elements situated on the red and blue sides of the bipolar major axis, all other components display symmetrical morphology.

\subsection{Constraints of the  model}
Low values of $\chi_{\text{red}}^2$ (between 1.0 and 2.0) indicate a close match between the simulated spectra and the observed ones. As shown in Table \ref{tab:T3_chi_square}, the estimated $\chi_{\text{red}}^2$ falls within this range for the majority of emission lines. However, some emission lines have $\chi_{\text{red}}^2$ red values that exceed the expected range. Notably, the most significant discrepancies were observed in the \ion{H}{$\alpha$} lines. As indicated in Table \ref{tab:T3_chi_square}, the $\chi^2$ values for each epoch are notably higher than expected. This discrepancy is likely due to the fact that the observed \ion{H}{$\alpha$} line in the spectrum appears as a broad rectangular shape, whereas \textsc{cloudy} typically generates a triangle-shaped emission line.
Consequently, we chose to exclude the \ion{H}{$\alpha$} line from the computation of $\chi_{\text{red}}^2$ in our model, as it becomes challenging to precisely match the line flux ratios to the observed values. Other lines contributing to higher $\chi^2$ values include \ion{H}{$\gamma$} on epochs 1 and 3, \ion{H}{$\zeta$} on epochs 1 and 5, \ion{He}{ii}~ (4686~\AA) on epochs 2 and 4, and \ion{N}{iii}~(4638~\AA) on epoch 1. These discrepancies may originated from the lines being suppressed either due to the model's inability to fully account for high optical depth or because the two-component model cannot adequately represent the intricate density structure of the nova ejecta.

Our model includes two distinct ejecta density components with higher and lower densities. This approach is considered more realistic than the one-component model, which assumes a uniform density across the ejecta shell. However, it may still fall short, as \textsc{cloudy} treats these components individually even if they are not truly distinct.
The \textsc{cloudy} model is unlikely to incorporate clumpy components embedded within a larger diffuse component, a common feature in novae. This limitation arises because \textsc{cloudy} is a one-dimensional code and is not suitable for describing such complex environments. In such cases, dynamical models would be more appropriate. While \textsc{cloudy} can perform dynamical simulations, they are nearly 1000 times slower than time-steady calculations \citep{2022ApJPandey}. The primary goal of this research is to develop a model that adequately explains the key characteristics of the shell ejecta from V1674 Her and its ionizing source. Currently, we have found that non-local thermal equilibrium (NLTE) time-independent snapshot models suffice. Using the employed modelling methods, we have successfully replicated a wide range of observational effects.

 The morpho-kinematic 3D model in this study suggests that the ejecta structure associated with \ion{H}{$\alpha$} consists of a non-spherical bipolar shape overlaid with an equatorial ring within the ejecta. However, for the purposes of simplicity in the cloudy modelling, we considered the geometry to be a homogeneous symmetric sphere, which does not entirely reflect reality. Temperature was held steady in our morpho-kinematic modelling since we observed that it had an insignificant impact on line profile fitting and 3D morphology. Consequently, temperature was not treated as a free parameter. \citet{2020MNRASHarvey} have also supported the notion that temperature has minimal effects on the structure of emission lines, particularly under conditions of higher densities.

\section{Summary}\label{sec7}
We have investigated the spectroscopic and photometric evolution of V1674 Her, the fastest-recorded nova. We employed the ionization code Cloudy to model the observed spectra and estimate the physical and chemical parameters of the nova system. Additionally, we determined the morphology and inclination angle of the nova system using the morpho-kinematical analysis tool \textsc{shape}. The main findings from our analysis are summarised below:
\begin{enumerate}[1.]	
\item The decline time scales, $t_{2}$ and $t_{3}$ are estimated to be $\sim$$0.91$ days and $1.94$ days, respectively, classifying it as an exceptionally rapid nova.
\item Utilising the MMRD relationship and a mean extinction of $A_V=1.705$ in the V band, we estimate the distance to the nova to be $\sim~4.97$ kpc.
\item We deduce the mass and radius of the WD to be $\sim~1.36~M_{\sun}$ and $\sim~0.15~R_{\earth}$, respectively, confirming the ONe-type nature of the WD and thus categorising the nova as a Neon-type. 
\item P Cygni absorption lines displayed an initial blue shift, evolving over time due to increasing velocity with distance.
\item The nova V1674 Her transformed from a \ion{Fe}{ii}-type to a \ion{He}{}/\ion{N}{}-type nova within ten days of the explosion.
\item The nova entered the nebular phase more rapidly than any recorded CNe, beginning on day 10 in the optical band. 
\item photoionization modelling of the observed spectra through \textsc{cloudy} provided estimates for the source's temperature and luminosity, as well as the ejecta's density and elemental composition (refer to results in Table \ref{tab:T2_results}). 
\item Our model, spanning one month after the outburst, unveiled an overabundance of \ion{He}{}, \ion{Ne}{}, \ion{O}{}, \ion{N}{}, and \ion{Fe}{} relative to solar abundances. After $\sim$17 days from the outburst, the \ion{Fe}{} abundance normalised to its solar value, while the neon abundance increased by a factor of 2, favouring the transition that occurred.
\item Based on our \textsc{cloudy} model, the ejected mass of V1674 is predicted to be $3.42\times 10^{-5}~M_{\sun}$, $4.23 \times 10^{-5}~M_{\sun}$, $4.35 \times 10^{-5}~M_{\sun}$, $4.85 \times 10^{-5}~M_{\sun}$, and $7.40 \times 10^{-5}~M_{\sun}$ for epochs 1, 2, 3, 4, and 5, respectively. These values align with theoretical predictions and prior findings for this nova.
\item The \ion{H}{$\alpha$} morphology of the nova ejecta during the initial month after the outburst exhibited asymmetry, featuring both bipolar and equatorial ring geometries. The presence of equatorial rings in the \ion{H}{$\alpha$} geometry suggests potential interactions of outflows with a secondary companion. We estimate the inclination angle of the ejecta geometry to be $\sim$$i = 67 \pm 1.5^{\circ}$.
\end{enumerate}

\section*{Acknowledgments}
We thank Prof. Robert Gehrz, the referee, for critically reading the manuscript and providing insightful comments, suggestions, and constructive questions, which significantly improved the paper.
We acknowledge the S. N. Bose National Centre for Basic Sciences and the World Academy of Science (TWAS) for their funding support.  We are grateful to F. Teyssier for coordinating the ARAS Eruptive Stars Section. Equally, we appreciate the ARAS observers for generously sharing their observations with the public. Special acknowledgments go to David Boyd, Pavol A. Dubovsky, Daniel Dejean, Robin Leadbeater, and Forrest Sims for their valuable spectroscopic observations. RP acknowledges the funding support from the Physical Research Laboratory, Ahmedabad, India. Similarly, PAD acknowledges the support from the Slovak Research and Development Agency under contract No. APVV-20-0148. Furthermore, we recognise and value the \textsc{AAVSO} database for providing open access to photometric data. We also extend our gratitude to Rahul Bandyopadhyay for engaging discussions regarding the \textsc{shape} software. GRH acknowledges the supports from Debre Berhan university, Debre Berhan, Ethiopia.

\section*{Data Availability}
In this paper, we utilised spectroscopic and photometric data sourced from the ARAS Database\footnote{ \url{https://aras-database.github.io/database/novae.html}.} and AAVSO\footnote{\url{https://www.aavso.org/LCGv2/} }, respectively. The spectroscopic data were employed to analyse the evolution of the ejecta (depicted in Fig.~\ref{fig:MRS} (a, b, \& c), and ~\ref{fig:LRS}) as well as for modelling purposes (illustrated in Fig.~\ref{fig:cloudy_best_fitting_model}). Conversely, the photometric data were utilised to assess the rate of brightness decline and generate the light curve (illustrated in Fig.~\ref{fig:lightcurvenher21}).


\bibliographystyle{mnras}
\bibliography{references} 







\bsp	
\label{lastpage}
\end{document}